\documentclass[usenatbib]{mn2e}
\usepackage{graphicx}
\usepackage{color}
\usepackage[a4paper, total={17.8cm, 24.0cm},centering]{geometry}
\graphicspath{{Figures/}}

\def\simgt{\hbox{\rlap{\raise 0.425ex\hbox{$>$}}\lower 0.65ex\hbox{$\sim$}}}
\def\simlt{\hbox{\rlap{\raise 0.425ex\hbox{$<$}}\lower 0.65ex\hbox{$\sim$}}}
\def\arcsec{^{\prime\prime}}

\def \pakin {PA$\rm{_{kin}}$}
\def \paphot {PA$\rm{_{phot}}$}
\def \twodfdr {2{\sc dfdr}}

\title[The SAMI Pilot Survey: Stellar Kinematics]{The SAMI Pilot Survey: Stellar Kinematics of Galaxies in Abell 85, 168 and 2399.}

\author[Fogarty et al.]
{L. M. R. Fogarty$^{1,2}$\thanks{l.fogarty@physics.usyd.edu.au}, N. Scott$^{1,2}$, M. S. Owers$^{3,4}$, S. M. Croom,$^{1,2}$,  K. Bekki$^{5}$,
\newauthor R. C. W. Houghton$^{6}$, J. van de Sande$^{1}$, F. D'Eugenio$^{7}$, G. N. Cecil$^{8}$, M. M. Colless$^{7}$,  
\newauthor J. Bland-Hawthorn$^{1}$, S. Brough$^{4}$, L. Cortese$^{9}$, R. L. Davies$^{6}$,  D. H. Jones$^{10,3}$,   
\newauthor M. Pracy$^{1}$,  J. T. Allen$^{1,2}$, J. J. Bryant$^{1,2,4}$, M. Goodwin$^{4}$, A. W. Green$^{4}$,
\newauthor I. S. Konstantopoulos$^{4}$, J. S. Lawrence$^{4}$,  N. P. F. Lorente$^{4}$, S. Richards$^{1,2,4}$, 
\newauthor R. G. Sharp$^{8}$ \\
$^{1}$ Sydney Institute for Astronomy, School of Physics, University of Sydney, NSW 2006, Australia.\\ 
$^{2}$ ARC Centre of Excellence for All-Sky Astrophysics (CAASTRO).\\
$^{3}$ Department of Physics and Astronomy, Macquarie University, NSW 2109, Australia. \\
$^{4}$ Australian Astronomical Observatory, PO Box 915, North Ryde, NSW 1670, Australia.\\
$^{5}$ ICRAR M468, The University of Western Australia, 35 Stirling Hwy, Crawley, Western Australia, 6009, Australia. \\
$^{6}$ Astrophysics, Department of Physics, University of Oxford, Denys Wilkinson Building, Keble Rd., Oxford, OX1 3RH, UK. \\
$^{7}$ Research School of Astronomy and Astrophysics, Australian National University, Canberra ACT 2611, Australia. \\
$^{8}$ Department of Physics and Astronomy University of North Carolina Chapel Hill, NC 27599 USA. \\
$^{9}$ Centre for Astrophysics and Supercomputing, Swinburne University of Technology, Hawthorn, VIC 3122, Australia.\\
$^{10}$ School of Physics, Monash University, Clayton, VIC 3800, Australia. \\
}

\begin{document}

\maketitle

\newcommand{\fmmm}[1]{\mbox{$#1$}}
\newcommand{\scnd}{\mbox{\fmmm{''}\hskip-0.3em .}}
\newcommand{\scnp}{\mbox{\fmmm{''}}}

\begin{abstract}

We present the SAMI Pilot Survey, consisting of integral field spectroscopy of 106 galaxies across three galaxy clusters, Abell 85, Abell 168 and Abell 2399. The galaxies were selected by absolute magnitude to have $M_r<-20.25$ mag. The survey, using the Sydney-AAO Multi-object Integral field spectrograph (SAMI), comprises observations of galaxies of all morphological types with 75\% of the sample being early-type galaxies (ETGs) and 25\% being late-type galaxies (LTGs). Stellar velocity and velocity dispersion maps are derived for all 106 galaxies in the sample.

The $\lambda_{R}$ parameter, a proxy for the specific stellar angular momentum, is calculated for each galaxy in the sample. We find a trend between $\lambda_{R}$ and galaxy concentration such that LTGs are less concentrated higher angular momentum systems, with the fast-rotating ETGs (FRs) more concentrated and lower in angular momentum. This suggests that some dynamical processes are involved in transforming LTGs to FRs, though a significant overlap between the $\lambda_{R}$ distributions of these classes of galaxies implies that this is just one piece of a more complicated picture.

We measure the kinematic misalignment angle, $\Psi$, for the ETGs in the sample, to probe the intrinsic shapes of the galaxies. We find the majority of FRs (83\%) to be aligned, consistent with them being oblate spheroids (i.e. disks). The slow rotating ETGs (SRs), on the other hand, are significantly more likely to show kinematic misalignment (only 38\% are aligned). This confirms previous results that SRs are likely to be mildly triaxial systems.

\end{abstract}


\section{Introduction}

Galaxies take many shapes, ranging from grand design spirals to smooth featureless ellipticals to highly irregular galaxies. Since shape is a readily apparent property, it has long been used to classify galaxies \citep[e.g.][]{Hubble1926, deVaucouleurs1959}. Early-type galaxies (ETGs) typically include those with featureless light profiles, the disky lenticulars (S0s) and the round ellipticals (Es). Conversely, the late-type galaxies (LTGs) are characterised by their spiral arm structure (tightly to loosely wound, Sa$-$Sd), sometimes seen with an accompanying bar (SBa$-$SBd).

The morphology-density relation \citep{Dressler1980, Houghton2015} shows that ETGs are found in higher relative numbers in regions of higher galaxy density, such as massive clusters. This trend is mostly driven by an increase in the number density of S0 galaxies at the expense of LTGs, prompting many studies of the transformation from LTGs to S0s. There are many mechanisms by which such transformations can occur, such as ram-pressure stripping, which removes gas from the galaxy \citep{GunnGott1972}; tidal encounters \citep{Icke1985, BC11}; truncation of the gas envelope, so that the remaining gas supply is rapidly consumed by SF without replenishment (known as strangulation) \citep{Larson1980, Bekki2002}; and minor mergers \citep{Bekki1998}. 

Different transformation mechanisms affect galaxies in different ways. However, morphology does not map to particular physical processes in a straightforward way. ETGs have been shown to exhibit a wide range of kinematic morphologies \citep{Krajnovic2011} and this can provide some clues as to their formation histories. The $\lambda_{R}$ parameter is a proxy for the specific stellar angular momentum of a galaxy \citep{Emsellem2007, Emsellem2011}, and is measured from 2D kinematic maps. Separating ETGs according to $\lambda_{R}$ yields two broad categories: fast and slow rotators (FRs/SRs). FRs and SRs have very different distributions in both angular momentum and intrinsic shape \citep{Cappellari2011b, Weijmans2014} and likely form differently. 

Establishing the different processes by which galaxies are transformed from LTGs to FRs, or from FRs or LTGs to SRs, is crucial to our understanding of galaxies. Large galaxy surveys using integral field spectroscopy (IFS), and with no morphological pre-selection, are needed to disentangle the different physical processes governing galaxy transformations. Observations of galaxy morphology are not enough - the addition of kinematic information provides a vital piece of the puzzle. In this paper we present the SAMI Pilot Survey, an IFS survey of 106 galaxies in three galaxy clusters. With this rich data set we explore evolutionary links between LTGs and FRs and the intrinsic structural differences between FRs and SRs.

The SAMI Pilot Survey is a precursor to the full SAMI Galaxy Survey\footnote{http://sami-survey.org/} \citep{Bryant2015} which will build a sample of $\sim3400$ galaxies over three years, each with IFS data. This large and complex data set will enable us to build on the work presented here and give a unique perspective to the way we think about galaxy evolution.

In Section \ref{sec:obs} we present our data and in Section \ref{sec:data_val} we validate our data reduction and analysis pipelines by comparing to previously published observations. In Section \ref{sec:der_params} we discuss some derived photometric parameters, such as measurements of the effective radius and ellipticity of the galaxies in our sample, which we use in our analysis. In Section \ref{sec:ltgs} we examine the distribution of angular momentum in our sample and in Section \ref{sec:kin_align} we discuss the kinematic alignment of our sample. We discuss our results in Section \ref{sec:discussion} and conclude in Section \ref{sec:conclusions}.

\section{Observations and Data}
\label{sec:obs}

\subsection{The SAMI Pilot Survey}
\label{sec:selection}

The Sydney--AAO Multi-object IFS (SAMI) is a fibre-fed multi-object IFS on the Anglo--Australian Telescope (AAT) \citep{Croom2012} capable of observing 13 galaxies at once. This has enabled an order of magnitude increase in survey speed compared to surveys using single object IFS. SAMI is based on lightly-fused fibre bundles called hexabundles \citep{JBHBryant2011, Bryant2011, Bryant2014}, with 13 hexabundles deployable across a one-degree field of view. Each SAMI hexabundle consists of 61 optical fibres arranged in a circular pattern with a mean fill factor of 73\%. Each fibre is 1.6$\arcsec$ in diameter and each hexabundle is $\sim15\arcsec$ in diameter on sky. SAMI is mounted on the triplet top-end of the AAT and feeds the AAOmega spectrograph \citep{Smith2004, Sharp2006}. AAOmega is a fully-configurable, double-beamed spectrograph. For general operations SAMI uses the low resolution 580V grating in the blue arm which delivers wavelength coverage in the range 3700--5700\AA\, at a resolution $R\sim1700$. In the red arm we use the medium resolution 1000R grating which delivers wavelength coverage in the range 6200--7300\AA\, at a resolution $R\sim4500$. This ensures the main stellar absorption features are covered in the blue while allowing precise emission line observations around H$\alpha$ in the red.

The SAMI Pilot Survey is a precursor to the SAMI Galaxy Survey and is focussed on galaxies in clusters. The cluster sample was selected from the X-ray cross-matched cluster catalogue of \citet{Wang2011arXiv}. Within this catalogue suitable clusters were constrained to lie at $\rm{z}<0.06$ and at $\rm{declination}<10^{\circ}$ so as to be observable from Siding Spring Observatory, with a total of seven clusters selected. Target galaxies were then selected from the NYU Value-Added Galaxy Catalogue \citep{Blanton2005} to be within a 1-degree radius of the cluster centre, and to be at $0.025<\rm{z}<0.085$. Initially galaxies within a redshift range of $\pm \rm{dz}=0.01$ of the cluster redshift were defined (conservatively) as cluster members and were the highest priority targets. This definition of membership was for target selection only and was later refined for the purposes of fulfilling our scientific aim \citep{Fogarty2014}.

All galaxies with $M_r<-20.25$\,mag were included in the full parent sample. Absolute magnitudes are from \citet{Blanton2005} and use a $\Lambda$CDM cosmology such that $\Omega_{M}=0.3$, $\Omega_{\Lambda}=0.7$ and $\rm{H_0=100\,kms^{-1}Mpc^{-1}}$. Cluster member galaxies meeting the conservative criteria above were prioritised according to absolute magnitude, with the brightest objects ($M_r<-21.68$\,mag) having highest priority, objects with $M_r<-20.5$\,mag having intermediate priority and objects with $-20.5$\,mag\,$<M_r<-20.25$\,mag having lower priority. Foreground and background galaxies around the cluster were similarly ranked by absolute magnitude (but all at a lower priority than the cluster galaxies). Of the selected clusters three were observed, Abell 85, Abell 168, and Abell 2399.

\subsection{Sampling and Completeness}
\label{sec:sampling}

The SAMI Pilot Survey was completed over two separate observing runs from 10th--13th September and 11th--16th October 2012. Not all 13 SAMI hexabundles were in use at this time due to hardware issues that were rectified after the October 2012 run. During the Pilot Survey fourteen fields, containing eight galaxies each, were observed. A ninth hexabundle was allocated to a secondary standard star for calibration purposes. In total 112 galaxies were observed. However, due to astrometric errors during the September 2012 run the data for six of these proved unusable. The final sample is therefore 106 galaxies (97 cluster members and 9 foreground/background objects) from three cluster fields of Abell 85, Abell 168 and Abell 2399. 

No morphological pre-selection was made and so all morphological types are included in our sample. The sample was classified visually using imaging into two morphological classes, ETGs and LTGs. For our purposes ETGs are objects with a smooth symmetrical light distribution without spiral structure or prominent dust lanes. All other galaxies are classified as LTGs (these are all spirals in our sample). A breakdown of the sample properties is given in Table \ref{tab:pilot_obs}. As the sample is dominated by cluster member galaxies the majority of galaxies in the sample are ETGs. Throughout this paper the terms ETG and LTG refer only to this visual morphological classification. 

At the sample selection and observation stage a simple estimate of cluster membership was made and so the final sample contains some foreground and background objects. Once the SAMI observations were complete a programme was undertaken to increase the spectroscopic completeness of the parent sample and to define cluster membership more precisely. The SAMI Cluster Redshift Survey (Owers et al. in prep.) used the 2dF $+$ AAOmega spectrograph on the AAT for 7 nights in September 2013. Redshifts were measured for $\sim1600$ cluster members within $R_{200}$, the cluster radius at which the density is 200 times the critical density of the Universe at that redshift, across 8 galaxy clusters including Abell 85, Abell 168 and Abell 2399. This enabled a robust measurement of cluster membership. The method used is described in detail in Owers et al. in prep., with an overview given in \citet{Fogarty2014}.

\begin{table*}
\centering
\begin{tabular}{|c|c|c|c|c|c|}
\hline
Cluster & Cluster & \multicolumn{2}{c}{ETGs} & \multicolumn{2}{c}{LTGs} \\
Name & Redshift & Members & Non-members & Members & Non-members \\
\hline
Abell 85 & 0.0549 & 28 & 0 & 2 & 0 \\
Abell 168 & 0.0451 & 12 & 0 & 11 & 1 \\
Abell 2399 & 0.0583 & 34 & 5 & 10 & 3 \\
\hline
\end{tabular}
\caption{A summary of the galaxies observed in the SAMI Pilot Survey. The majority ($\sim$75\%) of the sample comprises ETGs, with the remaining $\sim$25\% being LTGs. There are only 9 cluster non-members observed.}
\label{tab:pilot_obs}
\end{table*}

It is important that our sample of 97 cluster member galaxies is not biased such that we preferentially observe galaxies with a particular kinematic class \citep{Houghton2013}. This is particularly important when classifying the 74 ETGs in our cluster sample as FRs/SRs as SRs tend to be high mass (more luminous) galaxies, rounder than FRs (most SRs have ellipticity less than 0.4). To check for bias in our sample of 97 cluster members we compare the observed sample to the parent sample of cluster members from the original input catalogue, paying particular attention to potential bias in absolute magnitude and ellipticity distributions.

\begin{figure}
\centering
\includegraphics[width=8.5cm]{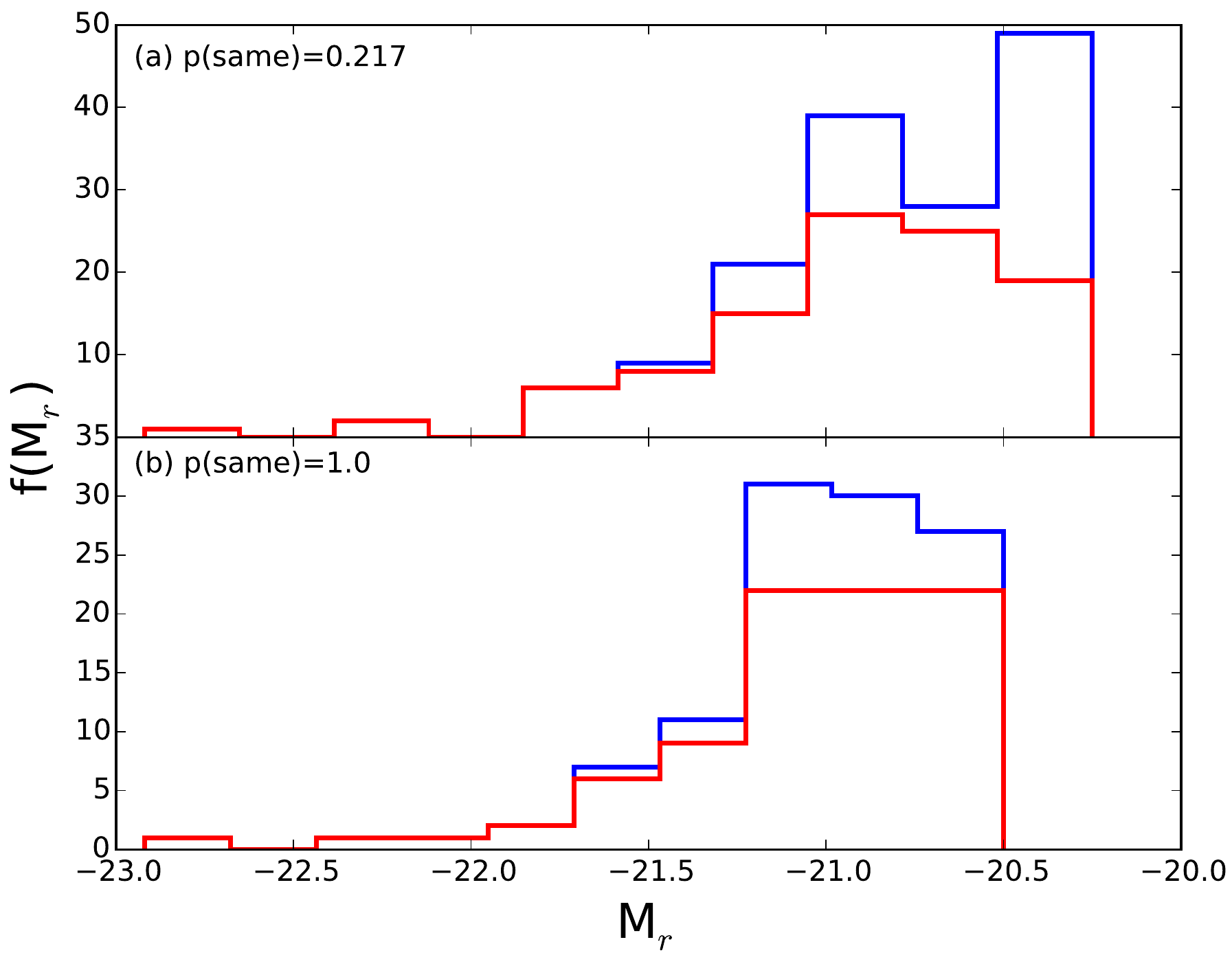}
\caption{The absolute $r$-band magnitude distribution of the parent cluster member sample selected from the NYU-VAGC \citep{Blanton2005} is shown in blue, with the observed SAMI Pilot cluster member galaxies shown in red. Panel (a) uses the parent sample with a cut-off of $M_r<-20.25$, including lower priority galaxies. Panel (b) shows just the high priority objects with $M_r<-20.5$. In both cases a K--S test convincingly shows that there is no measurable bias in our observed sample.}
\label{fig:lum_comp}
\end{figure}

\begin{figure}
\centering
\includegraphics[width=8.5cm]{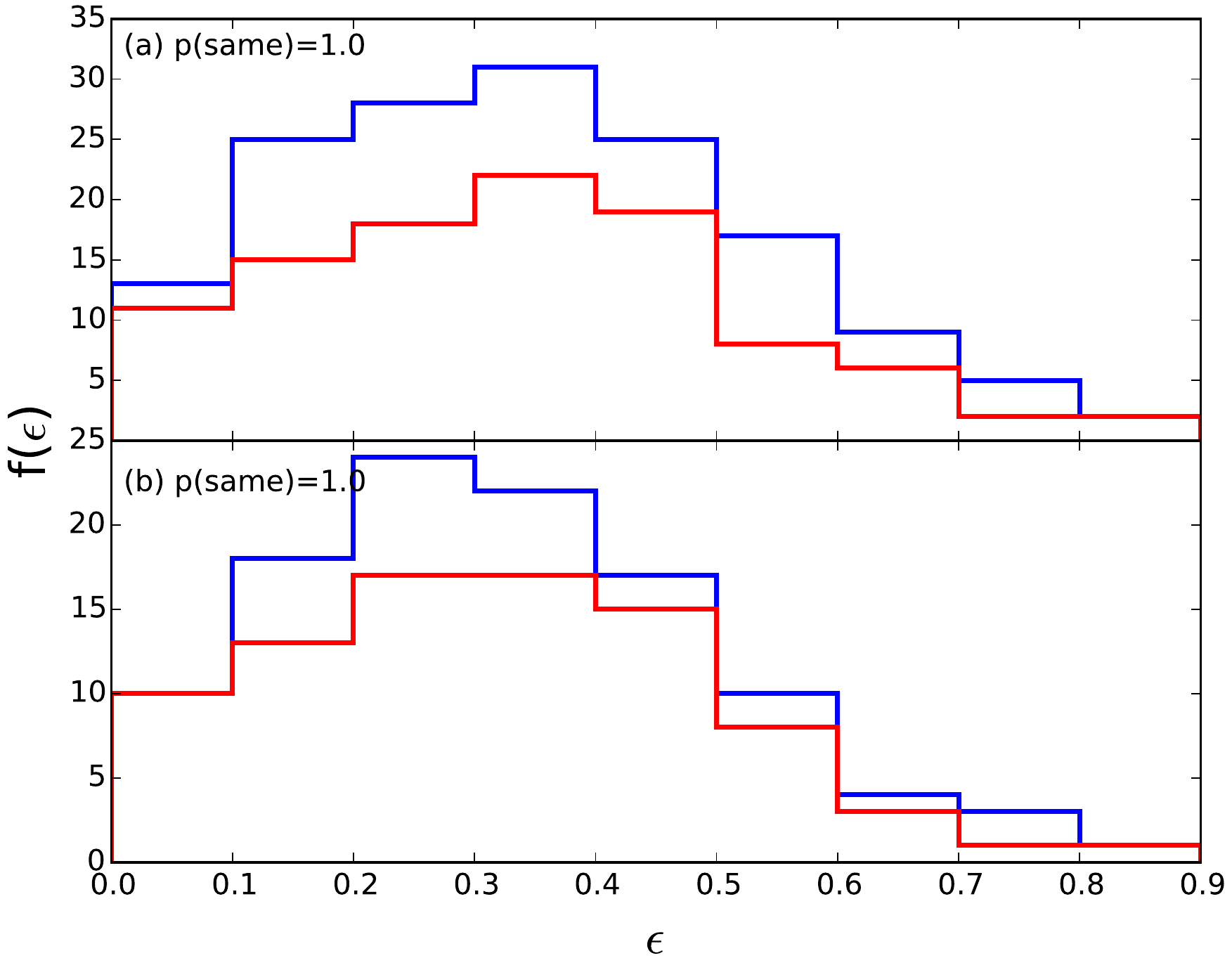}
\caption{The ellipticity distribution of the parent cluster member sample selected from the NYU-VAGC \citep{Blanton2005} is shown in blue, with the observed SAMI Pilot cluster member galaxies shown in red. Panels are as for Figure \ref{fig:lum_comp} with the full sample shown in panel (a) and the high priority bright sample shown in panel (b). Again, there is no measurable bias in our observed sample.}
\label{fig:ellip_comp}
\end{figure}

The distribution of $r-$band absolute magnitude ($M_r$) is shown in Figure \ref{fig:lum_comp}. Two comparisons are made, first including the low priority targets (those with $-20.5$\,mag\,$<M_r<-20.25$\,mag) and second considering only the high priority targets (i.e. with $M_r<-20.5$\,mag). In the first case a Kolmogorov--Smirnov (K--S) test yields a p-value of 0.217 meaning that we cannot reject the null hypothesis that the two distributions are drawn from the same parent sample. In the case of the high priority targets the p-value is 1.0. It is not surprising that the p-value should be lower in the case where we include the lower priority targets. However, in both cases it is clear that there is no measurable bias in our observed sample. The distribution of ellipticities is shown in Figure \ref{fig:ellip_comp} with the panels as per Figure \ref{fig:lum_comp}. For both cases the p-values is 1.0 and this convincingly shows that our sample is not biased in ellipticity.

\subsection{Test Galaxy Observations}

\begin{table*}
\centering
\begin{tabular}{|c|c|c|c|c|c|}
\hline
Galaxy & R.A. (J2000) & Dec. (J2000) & Distance & Date & Exposure \\
 & (degrees) & (degrees) & (Mpc) & Observed & Time \\
\hline
NGC1289 & 49.707542 & -1.973333 & 38.4 & 14th October 2012 & 2\,x\,900s \\ 
NGC1665 & 72.071142 & -5.427607 & 37.5 & 15th October 2012 & 3\,x\,900s \\ 
\hline
\end{tabular}
\caption{The two galaxies selected from the ATLAS$^{\rm{3D}}$ survey and observed with SAMI in October 2012 as part of data validation tests. Distances are taken from \citet{Cappellari2011a}.}
\label{tab:test_gals}
\end{table*}

To validate our data reduction and analysis procedures we compare SAMI observations with already public IFS data. The ATLAS$^{\rm{3D}}$ survey \citep{Cappellari2011a} provides a useful sample for comparison. The stellar kinematics for the 260 ETGs observed for the ATLAS$^{\rm{3D}}$ survey and presented in \citet{Krajnovic2011} are now public.

During the SAMI Pilot Survey run in October 2012 we observed two galaxies from the ATLAS$^{\rm{3D}}$ sample, NGC1289 and NGC1665. These galaxies are not typical SAMI Pilot Survey or SAMI Galaxy Survey targets, being much more local and therefore larger on the sky. Both galaxies significantly overfill the SAMI hexabundle field of view. Despite this, these observations do give us the opportunity to test whether we can reproduce previously known results with our instrument, observation procedure, and data reduction and analysis tools. Details of our test galaxy observations are given in Table \ref{tab:test_gals}.

Both NGC1289 and NGC1665 were observed using a single hexabundle in SAMI, with the other hexabundles observing blank sky. This is not the normal operating procedure for SAMI and led to some subtle differences in the data processing. These are described in detail in Section \ref{sec:samidr}.

\subsection{SAMI Data Reduction}
\label{sec:samidr}

\begin{figure}
\centering
\includegraphics[width=8cm]{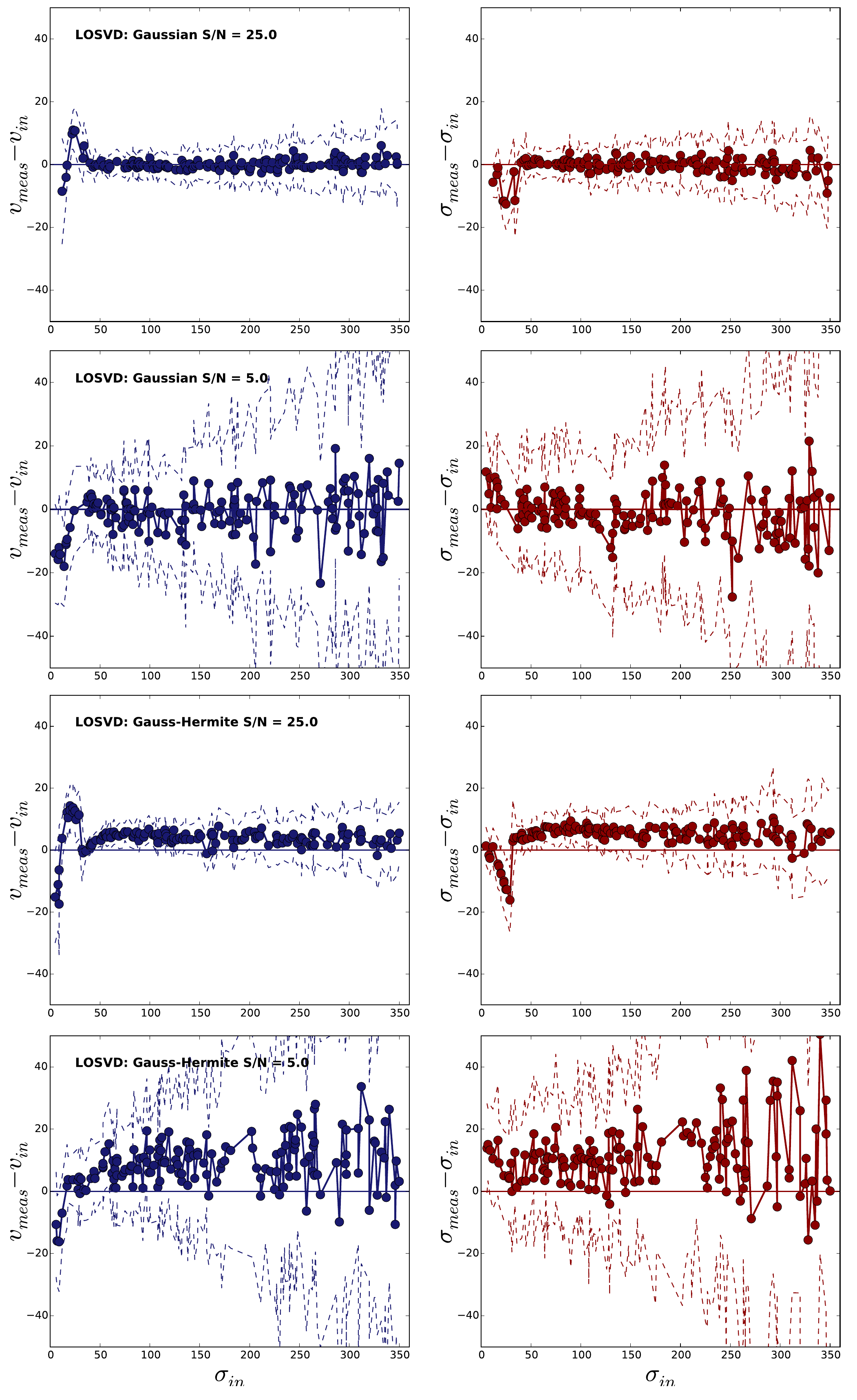}
\caption{Results from simulated LOSVD fitting with pPXF. The top four panels show the simulations using a Gaussian LOSVD as input, for a spectrum with S/N=25 (top row) and one with S/N=5 (second row). The bottom four panels show the simulations using a Gauss-Hermite LOSVD, for a spectrum with S/N=25 (third row) and one with S/N=5 (bottom row). In all cases the blue points show the difference between the input velocity and the measured velocity as a function of input velocity dispersion. The red points show the  difference between the input velocity dispersion and the measured velocity dispersion. The solid horizontal line indicated zero and the dashed lines show the one-sigma uncertainties.}
\label{fig:ppxf_SN}
\end{figure}

The data reduction procedure for the SAMI Pilot survey is similar to that for the full SAMI Galaxy Survey \citep{Allen2015, Sharp2015}, but differs in a few small ways. This is due mainly to the nature of the SAMI Pilot Survey observations which used a less uniform dither pattern (i.e. different for different fields). The SAMI Pilot Survey data are of lower quality than the SAMI Galaxy Survey data as an instrument upgrade was performed between the two surveys, significantly improving the throughput of the instrument [see \citet{Bryant2015} for details]. Here we briefly describe the  data reduction procedure and the areas where the SAMI Pilot Survey data and our test data are treated differently.

The raw SAMI data are initially processed using the \twodfdr\footnote{http://www.aao.gov.au/science/software/2dfdr} package \citep{SharpBirchall2010}. For each raw frame, the output from \twodfdr\, is a reduced row stacked spectrum (RSS) frame with all 819 SAMI fibre spectra bias subtracted, flat-fielded, wavelength calibrated and sky subtracted. The remaining reduction steps are performed using the custom-written SAMI python package \citep{Allen2014}.  After the RSS data are produced, a spectrophotometic standard star is used to flux calibrate data. The variance for each fibre spectrum is correctly propagated through these steps. 

As the fill factor of the SAMI hexabundles is not 100\%, the standard observing procedure is to dither the observations, filling in the gaps between fibres and gaining spatial resolution \citep{Sharp2015}. Each dither yields a single RSS file, all of which must then be combined to form a uniformly sampled data cube for each galaxy. For the SAMI Pilot Survey the dither pattern used was non-standard. A different set of parameters to those used for the SAMI Galaxy Survey is therefore adopted when resampling the SAMI Pilot Survey data cubes. Similarly for the test galaxy data cubes, a unique set of resampling parameters are used, suitable for these data only.

To resample the SAMI RSS frames to uniform data cubes we use an algorithm similar to Drizzle \citep{FruchterHook2002}. The procedure, fully described in \citet{Sharp2015}, is based on two parameters: a drop factor and output grid sampling. The drop factor is used to ``shrink'' the round footprint of the input fibres, conserving flux density. The new shrunk footprint is drizzled onto an output grid of square spaxels, the size of which is given by the output grid sampling. The SAMI Pilot Survey uses a drop factor of 0.75 and an output grid sampling of 0.5$\arcsec$ per spaxel. For the test galaxies the observation procedure is very tailored. Notably there are only 2 and 3 dithers for NGC1289 and NGC1665 respectively, and so a drop factor of 1.0 is chosen to minimise gaps in coverage. Because of the large drop size the output grid sampling is also chosen to be large, 1$\arcsec$ per spaxel to minimise covariance between spatial pixels.

When resampling the SAMI RSS frames onto regularly gridded data cubes it is important to correctly align the individual dithers with one another. This step is performed using fits to the secondary standard star which is observed simultaneously with all galaxy fields in the SAMI Pilot Survey (see \citet{Allen2015} for details). However, our test data have no such secondary star. Instead the alignment between RSS frames is found by fitting the position of the galaxy in each frame, using a two-dimensional Gaussian profile, and registering them according to the fit positions. Since both test galaxies are very bright and are smooth regular objects this procedure works well. Atmospheric dispersion is neglected for the test galaxy observations, but this does not impact our results.

\subsection{Stellar Kinematics of galaxies in Abell 85, Abell 168 and Abell 2399}
\label{sec:stellarkin}

\begin{figure*}
\centering
\includegraphics[width=16cm]{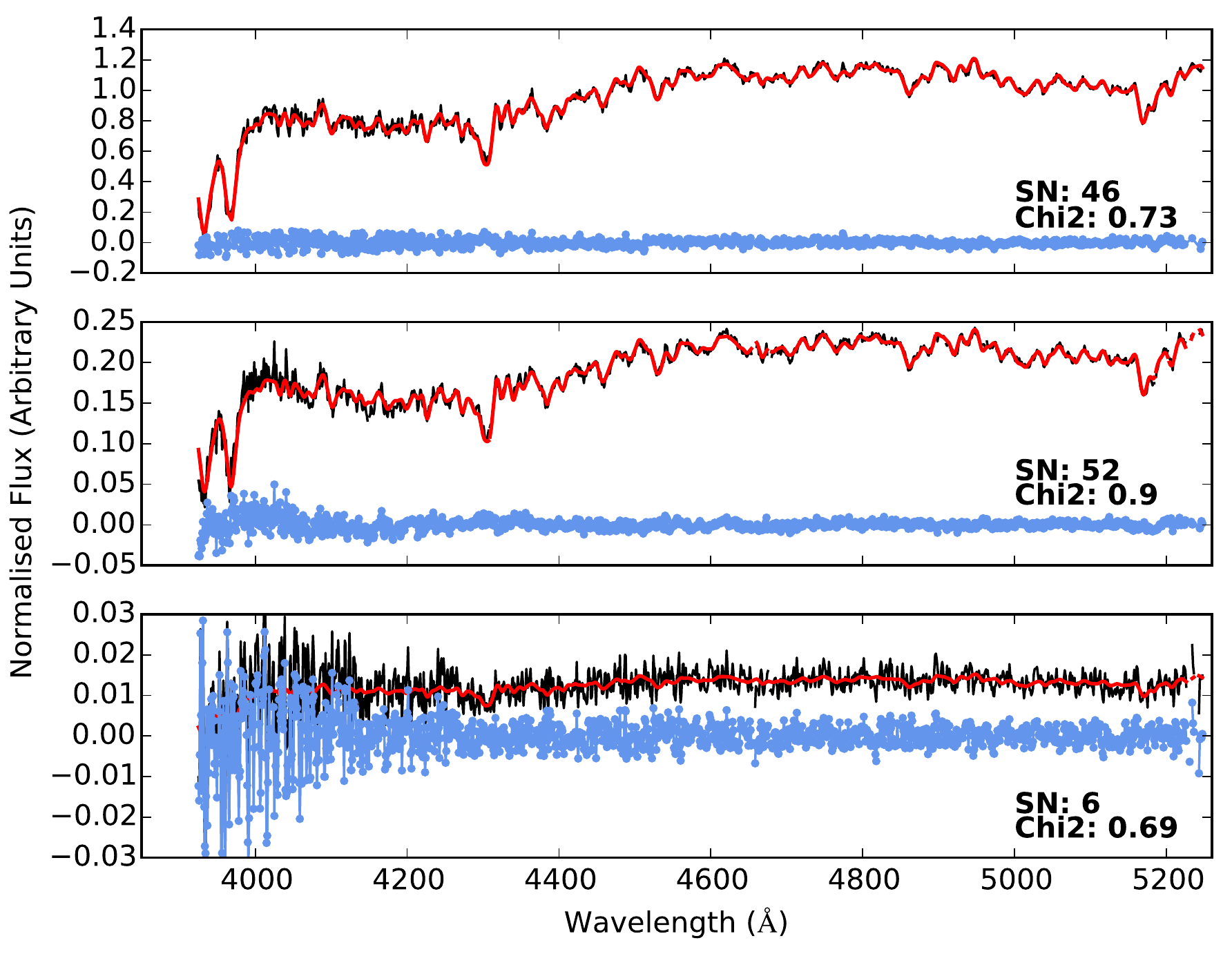}
\caption{Three SAMI spectra with pPXF fits are shown for the galaxy J004001.68-095252.5. In the top panel the integrated aperture spectrum (from a \rm{6}$\arcsec$ radius circular aperture) is shown in black, with the pPXF fit in red and the residuals in blue. The centre panel shows a single spaxel spectrum from the centre of the galaxy, with high S/N. The bottom panel shows a lower S/N spectrum from the outskirts of the galaxy, with S/N close to our chosen threshold of 5. The low S/N spectrum shows a good fit above 4100\,\AA\, but with increased noise below.}
\label{fig:ppxf_fits}
\end{figure*}

We use the penalised pixel-fitting routine (pPXF) developed by \citet{CappellariEmsellem2004} to fit stellar templates to the SAMI Pilot Survey galaxies. The procedure we followed is described below.

For each galaxy a spectrum is extracted from a central circular aperture with a 6$''$ radius. This spectrum is fit using pPXF and the 985 MILES stellar templates \citep{SanchezBlazquez2006} (with a fourth order additive polynomial) determining the best fit set of templates for that spectrum. The weighted and combined best fit template is then fit (again with a fourth order additive polynomial) to every spaxel in the data cube that meets the criterion of S/N$\geq5$ in the continuum. The S/N is measured across a narrow wavelength range ($\sim200$\,\AA) in a relatively flat part of the spectrum defined in pixel space and corresponding to between 4716\,\AA\, and 4928\,\AA\, for the lowest redshift object in our sample and 4913\,\AA\, and 5134\,\AA\, for the most distant. In all fits the emission lines are clipped using the \textit{clean} option in pPXF. This works well for the SAMI Pilot Survey galaxies as none contain broad emission lines with significant wings.

The S/N threshold was chosen to balance the need for spatial coverage in the outskirts of the galaxies in the sample with the need for high quality kinematic measurements. If the S/N threshold is too low we risk measuring kinematics that are systematically biased. To test this we construct a set of spectra with fixed S/N values and convolved the spectra with a known LOSVD, parametrised by either a Gaussian or by a Gauss-Hermite series, with four moments \citep{vanderMarelFranx1993, Gerhard1993}:

\begin{equation}
L_{gh}(v)=\frac{1}{\sigma \sqrt{2\pi}} e^{-\frac{(v-V)^2}{2\sigma^2}} \left[1+\sum_{m=3}^{4}h_{m}H_{m}(y)\right]
\end{equation}
 
Both the Gaussian and Gauss-Hermite LOSVDs are constructed with no velocity shift, for simplicity. The latter uses higher order moments $h_{3}=h_{4}=0.1$. For both versions of the LOSVD the input velocity dispersion is varied between $5-350\,\rm{km\,s^{-1}}$. A template spectrum is constructed by fitting the central aperture spectrum of one of the SAMI Pilot Survey galaxies (J004001.68-095252.5) using the MILES spectral library. The MILES spectra chosen by pPXF are combined to create a single infinite S/N spectrum with no intrinsic stellar velocity dispersion. Gaussian noise is then added to create spectra with the desired S/N, in this case we tested S/N values of 25 and 5. We then measure the kinematics of the resulting spectra in the same way as for the SAMI Pilot Survey data, fitting for all four moments ($V$, $\sigma$, $h_{3}$ and $h_{4}$) using the automatic bias setting in pPXF \citep{CappellariEmsellem2004}. 

The results are shown in Figure \ref{fig:ppxf_SN} for the Gaussian LOSVD and an input S/N of 25 and 5 (top four panels). At S/N=25 we do not find any systematic offsets in the recovered velocity and velocity dispersion down to an input dispersion of about $30\,\rm{km\,s^{-1}}$. At a S/N of 5 the $V$ and $\sigma$ values are also recovered without systematic errors, though the random errors are larger when compared with the values recovered from the spectrum with S/N=25. 

If we assume a Gauss-Hermite LOSVD (Figure \ref{fig:ppxf_SN}, bottom four panels) the recovered values show a small systematic bias in $V$ and $\sigma$ on the order of $\sim5-10\,\rm{km\,s^{-1}}$. This is due to the pPXF penalising bias setting not being perfectly tuned to the data. However, this is a small effect and our simulations show we are dominated by random errors, not systematic errors. For our purposes we assume that the LOSVD is Gaussian and accept that if this assumption is not always correct we introduce a small error in the measurements.

The central aperture radius, which is used to determine the optimal template for each galaxy, was chosen to account for strong stellar population gradients present in some of the LTGs in the sample. \citet{Fogarty2014} used a 2$\arcsec$ radius central aperture spectrum, which may not represent the entire galaxy. This method is fine for ETGs, which tend to display a weak stellar population gradient, if any. To investigate whether template mismatch affects the kinematics for our sample we test two central apertures to determine the best fit template, a central 2$\arcsec$ radius aperture and a 6$\arcsec$ radius aperture covering the whole galaxy. We make this comparison for all 106 galaxies, comparing the resulting velocity dispersion maps to search for a systematic bias. In the majority of cases the kinematic maps are the same within the errors. However, in eight of the LTGs (30\%) a systematic bias in $\sigma$ is seen. The offset is small ($<10\,\rm{km\,s^{-1}}$) and within the random error for most galaxies, but for those with the strongest radial colour gradients a systematic bias of greater than $20\,\rm{km\,s^{-1}}$ is seen. Template mismatch is a measurable problem for about half the LTGs in our sample. Using a large aperture to determine the best fit templates mediates this issue somewhat. For this reason we use the 6$\arcsec$ radius aperture spectrum to determine the optimal template for all galaxies in the sample when extracting kinematics. This has a small impact on values derived from the kinematic maps, such at the kinematic PA and $\lambda_{R}$. The latter values are therefore different to those reported in \citet{Fogarty2014} for the ETGs. This is briefly discussed in Section \ref{sec:lr}


Some typical pPXF fits are shown in Figure \ref{fig:ppxf_fits} for the ETG J004001.68-095252.5. The top panel shows the fit to the central spectrum, extracted in a 6$\arcsec$ radius circular aperture centred on the galaxy. The S/N in this spectrum is 46 and the fit is good, with a value of $\chi^{2}_{red}=0.73$. The centre and bottom panels of Figure \ref{fig:ppxf_fits} show single spaxel spectra from the data cube of the same galaxy. The centre panel shows a high S/N spectrum from the central part of the galaxy, where the pPXF fit is still very good, with a $\chi^2_{red}=0.9$. The lower panel shows a spectrum near to our S/N cut-off criterion. The fit to this spectrum has a lower value of $\chi^2_{red}=0.69$. This spectrum is much noisier than the others, especially at wavelengths shorter than $4100\rm{\AA}$ where there is essentially no signal.

The same procedure is used to extract stellar kinematics for NGC1289 and NGC1665.

\section{Data Validation}
\label{sec:data_val}

\begin{figure}
\centering
\includegraphics[width=8.5cm]{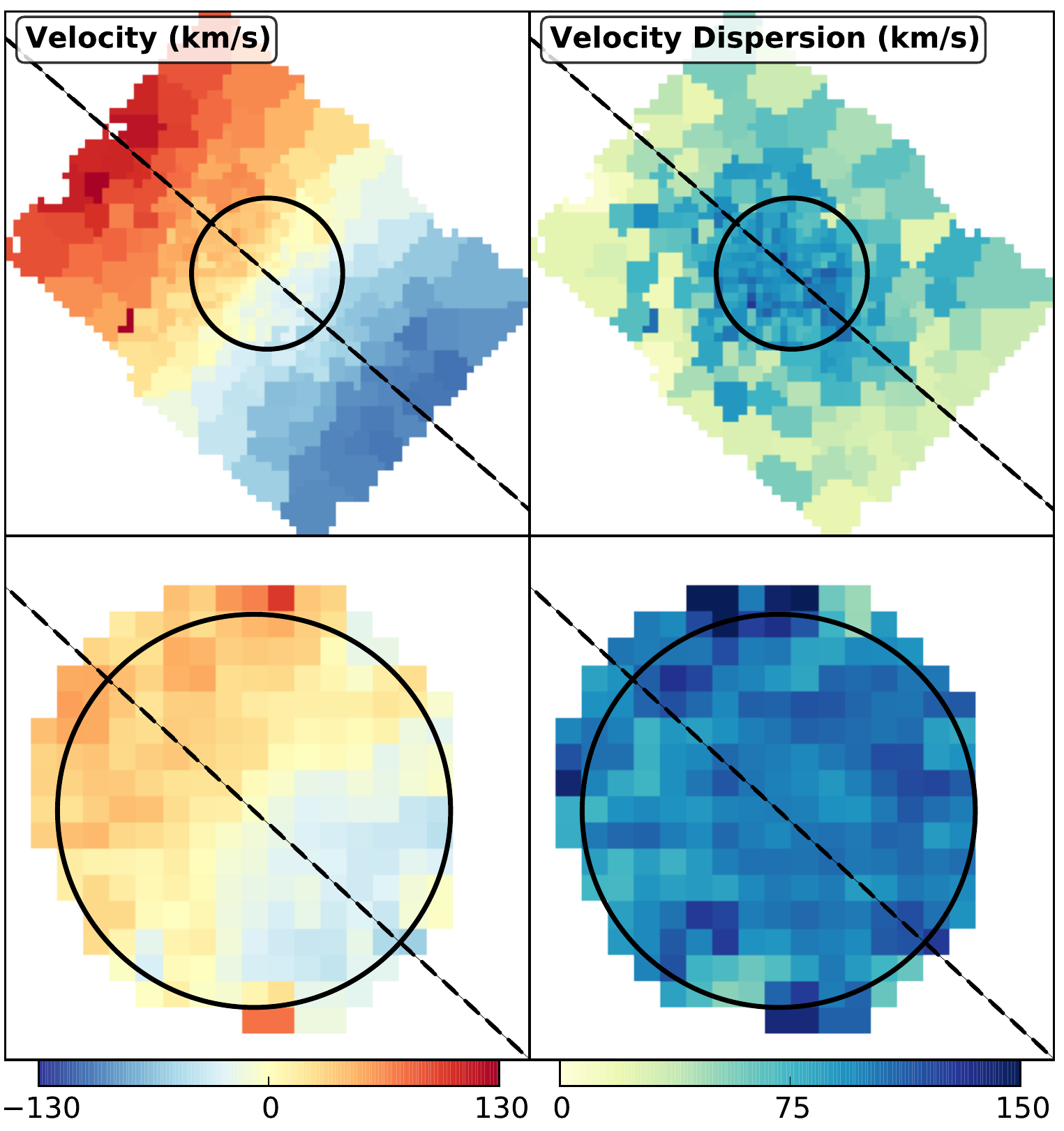}
\caption{The stellar kinematics maps of NGC1665 from the ATLAS$^{\rm{3D}}$ survey (top row) and from SAMI (bottom row). In each panel North is up and East to the left. The left hand columns shows the velocity maps, scaled to $\pm130\,\rm{km\,s^{-1}}$ and the right-hand column shows the velocity dispersion, scaled to $0 - 150\,\rm{km\,s^{-1}}$. The circle shows the SAMI hexabundle field of view, with a diameter of $\sim15''$ and the dashed line shows the PA along which the profiles in Figure \ref{fig:ngc1665_profiles} are extracted. The structure in the SAMI maps matches that in the ATLAS$^{\rm{3D}}$ maps.}
\label{fig:ngc1665_maps}
\end{figure}

\begin{figure}
\centering
\includegraphics[width=8.5cm]{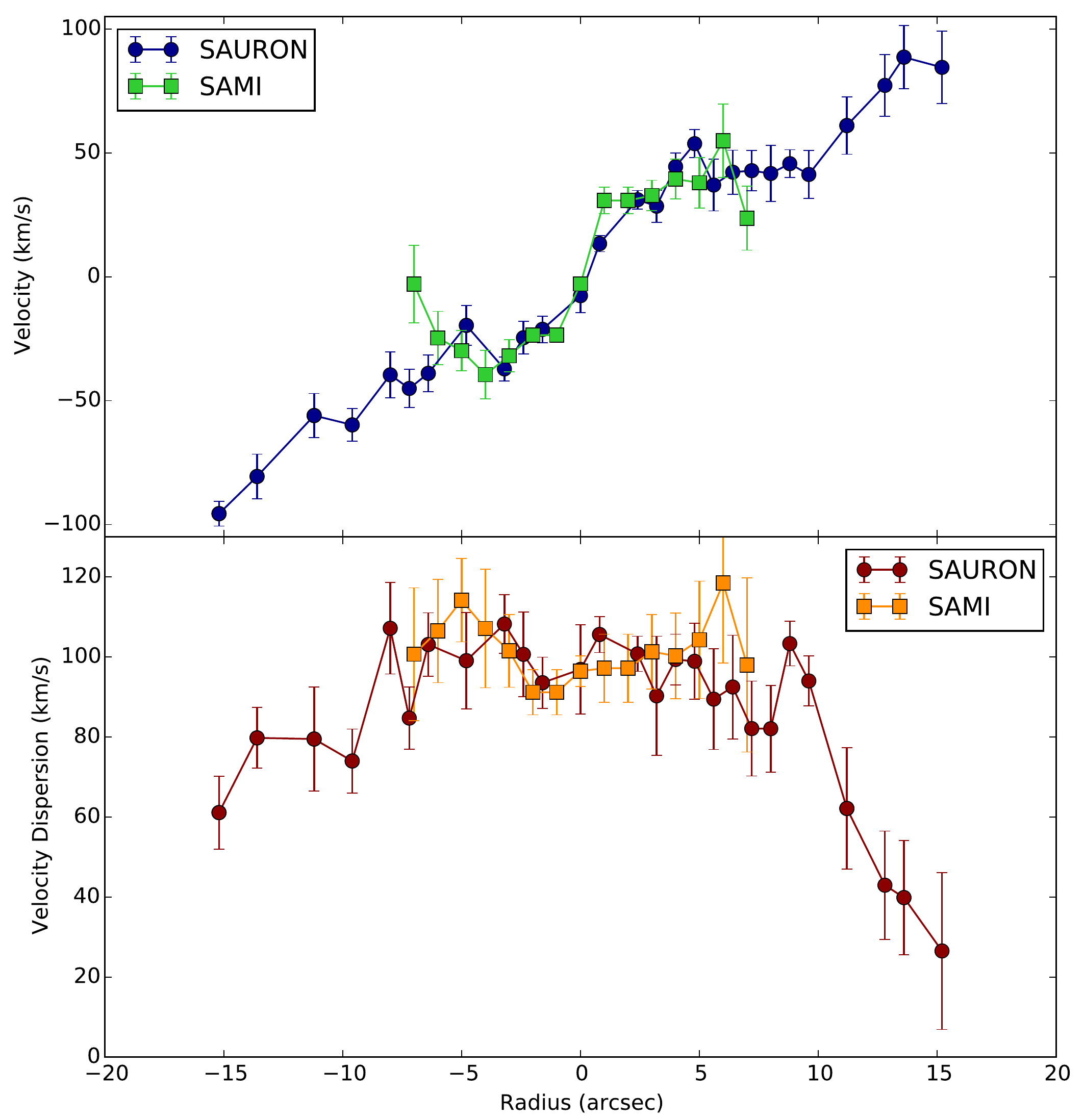}
\caption{Radial profiles of the kinematics of NGC1665 from the ATLAS$^{\rm{3D}}$ survey and from SAMI. The top panel shows the rotation curve derived from ATLAS$^{\rm{3D}}$ in dark blue circles and from SAMI in light green squares. The bottom panel shows the velocity dispersion profile derived from ATLAS$^{\rm{3D}}$ in dark red circles and from SAMI in orange squares.}
\label{fig:ngc1665_profiles}
\end{figure}

As with any new instrument it is extremely important that we validate our observations by studying previously observed objects with published measurements. To do this we observed two test galaxies drawn from the ATLAS$^{\rm{3D}}$ sample \citep{Cappellari2011a}. These are NGC1665 and NGC1289. The SAMI observations of these galaxies are compared to publicly available data from ATLAS$^{\rm{3D}}$.

Two nearby ETGs selected from the ATLAS$^{\rm{3D}}$ survey sample were observed with a single SAMI hexabundle in October 2012. \citet{Emsellem2011} classify NGC1665 as a FR with NGC1289 classified as a SR. Details of the SAMI observations for each galaxy are given in Table \ref{tab:test_gals}. These galaxies are not typical SAMI targets, being much larger on sky. Thus we are only able to probe the inner parts of these objects as compared to the broader view afforded by the 33$\arcsec$ x 41$\arcsec$ SAURON IFS observations \citep{Bacon2001} used for the ATLAS$^{\rm{3D}}$ survey. Nonetheless a valuable comparison of the galaxy centres can be made.

\subsection{NGC1665}

The kinematic maps for NGC1665 are shown in Figure \ref{fig:ngc1665_maps}. Data from the ATLAS$^{\rm{3D}}$ survey is shown on the top row and our new SAMI observations are on the bottom row. The velocity maps are on the left, with velocity dispersion on the right. The colour scaling is the same for corresponding ATLAS$^{\rm{3D}}$ and SAMI maps. A qualitative examination of the maps shows that the structure in the SAMI and ATLAS$^{\rm{3D}}$ data is very similar. 

We also produce radial profiles of velocity and velocity dispersion from the ATLAS$^{\rm{3D}}$ and SAMI maps, shown in Figure \ref{fig:ngc1665_profiles}. The profiles are extracted by plotting every spaxel which intersects a line defined by the photometric position angle of the galaxy, as given in \citet{Cappellari2011a}. The line is shown as a dashed black line in Figure \ref{fig:ngc1665_maps}. The rotation curves are shown in the top panel. Within the SAMI field of view ($\sim\pm7.5\arcsec$) the ATLAS$^{\rm{3D}}$ and SAMI curves agree well, within the errors, with both curves reaching $\pm50\,\rm{km\,s^{-1}}$. The same is true for the velocity dispersion profiles shown in the bottom panel of Figure \ref{fig:ngc1665_profiles}. In this case the SAMI observations do not extend to the fall off seen by ATLAS$^{\rm{3D}}$ at $\sim\pm10\arcsec$. However, in the central region both SAMI and ATLAS$^{\rm{3D}}$ see a flat velocity dispersion profile with a value of $\sim100\,\rm{km\,s^{-1}}$.

\subsection{NGC 1289}

\begin{figure}
\centering
\includegraphics[width=8.5cm]{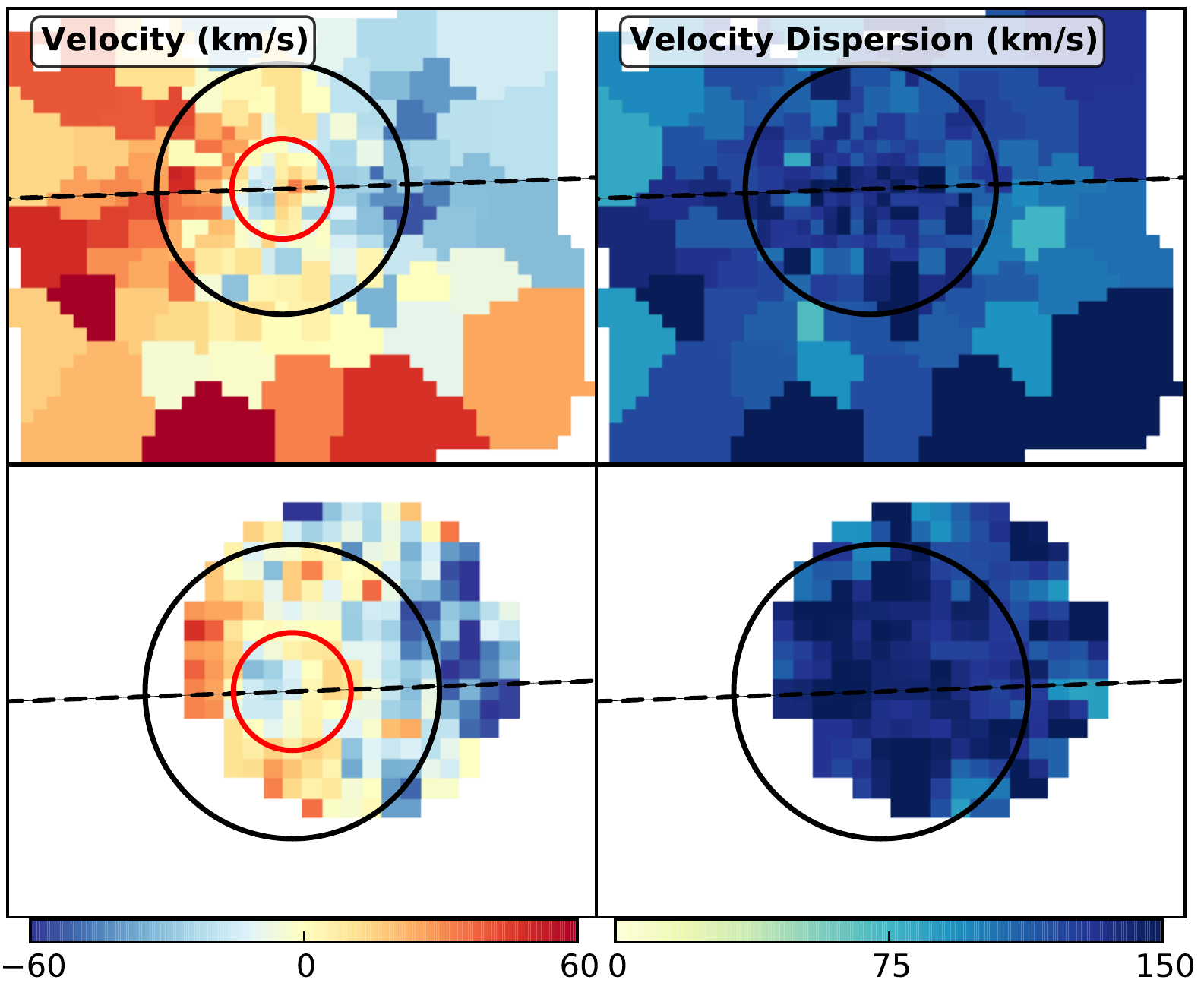}
\caption{The stellar kinematics maps of NGC1289 as measured by the ATLAS$^{\rm{3D}}$ survey (top row) and by SAMI (bottom row). In each panel North is up and East to the left. The left hand columns shows the velocity maps, scaled to $\pm60\,\rm{km\,s^{-1}}$ and the right-hand column shows the velocity dispersion, scaled to $0 - 150\,\rm{km\,s^{-1}}$. The black circle shows the SAMI hexabundle field of view, centred on the galaxy (though in this case the observation was not well centred), with a diameter of $\sim15''$ and the dashed line shows the PA along which the profiles in Figure \ref{fig:ngc1289_profiles} are extracted. The red circle highlights the kinematically decoupled core in this galaxy. As seen for NGC1665 the structure in the SAMI maps matches that in the ATLAS$^{\rm{3D}}$ maps.}
\label{fig:ngc1289_maps}
\end{figure}

The kinematic maps for NGC1289 are shown in Figure \ref{fig:ngc1289_maps}. Note that for this object the galaxy was not well-centred in the SAMI hexabundle. As for NGC1665 the structure in the ATLAS$^{\rm{3D}}$ and SAMI maps is  very similar. In particular this galaxy possesses a kinematically decoupled central component \citep[KDC;][]{Krajnovic2011} which is clearly seen in both the ATLAS$^{\rm{3D}}$ and SAMI velocity maps. The KDC is indicated by the red circle in the velocity maps, shown in the left-hand column of Figure \ref{fig:ngc1289_maps}.

\begin{figure}
\centering
\includegraphics[width=8.5cm]{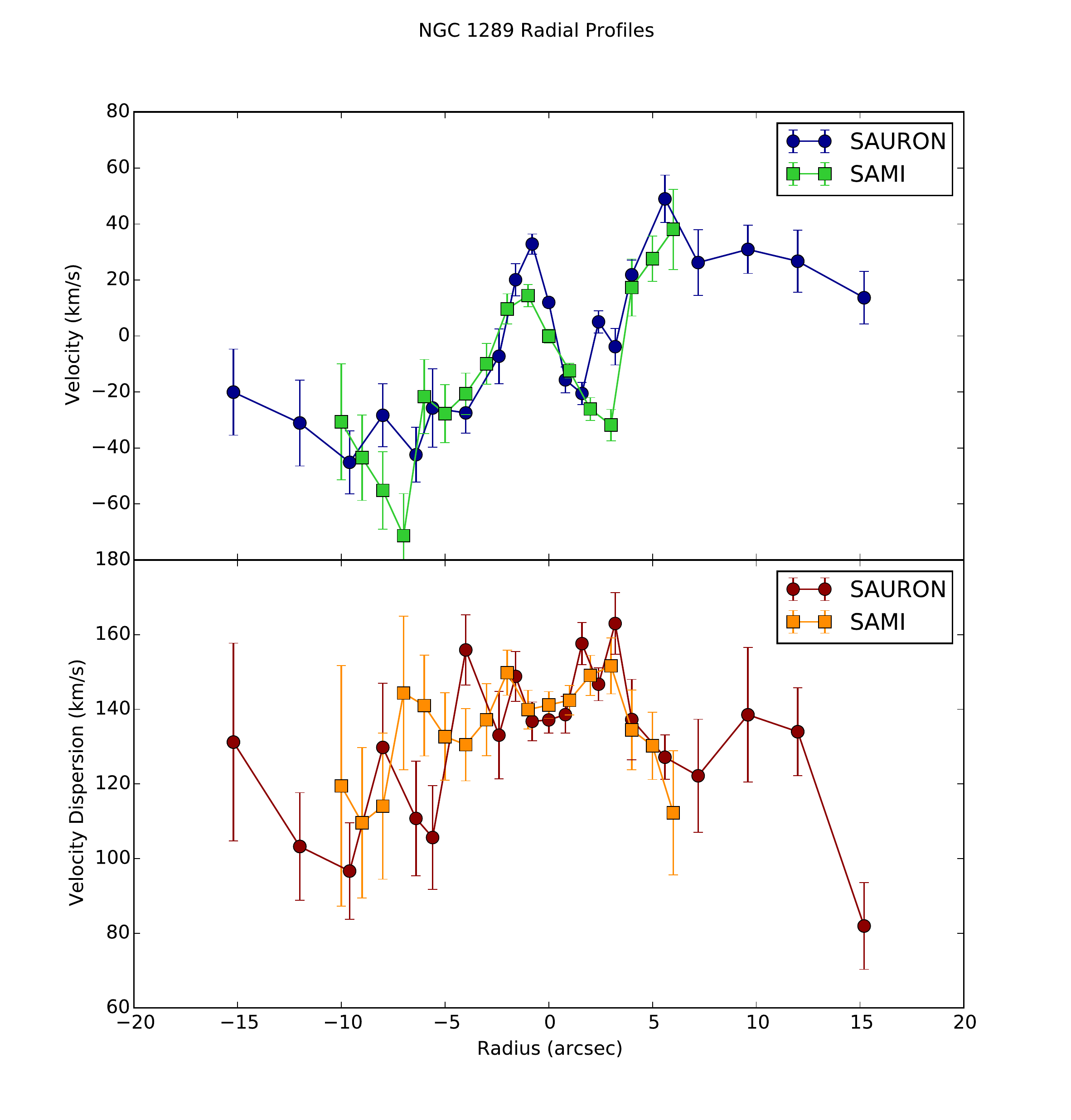}
\caption{Radial profiles of the kinematics of NGC1289 as measured by ATLAS$^{\rm{3D}}$ and by SAMI. The curves are as for Figure \ref{fig:ngc1665_profiles}.}
\label{fig:ngc1289_profiles}
\end{figure}

We extract radial profiles in the same way as for NGC1665, along a line defined by the PA of the galaxy, shown as a dashed black line in Figure \ref{fig:ngc1289_maps}. The profiles are shown in Figure \ref{fig:ngc1289_profiles}. The presence of the KDC is apparent in both rotation curves in the top panel of Figure \ref{fig:ngc1289_profiles}. The characteristic shape of the inner region of the curve, with a sharp rise and fall, superimposed on a more regular rotation curve clearly indicates a KDC. There is excellent agreement between the ATLAS$^{\rm{3D}}$ and SAMI data. The velocity dispersion profile also shows good agreement, and in this case SAMI does probe far enough in radius to start to see the same drop off in velocity dispersion seen by ATLAS$^{\rm{3D}}$.

The excellent agreement between the new SAMI observations and the ATLAS$^{\rm{3D}}$ data shows that our observing strategy and data analysis techniques can replicate results from the literature.

\section{Derived Parameters}
\label{sec:der_params}

\subsection{Photometric Parameters}
\label{sec:phot_pars}

We derive photometric parameters for each galaxy in our sample using theSloan Digital Sky Survey (SDSS) $r$-band images \citep{SDSSDR8_2011}. For the analysis presented here we are interested in the effective radius (R$_e$), and the ellipticity ($\epsilon$) and photometric position angle (PA$\rm{_{phot}}$) measured at the effective radius for each galaxy.

The approach we follow is described in \citet{Fogarty2014}. Briefly, we use a Multi Gaussian Expansion \citep[MGE;][]{Emsellem1994} of the surface brightness profile to derive an accurate measurement of the effective radius of each galaxy. Separately, profiles of the ellipticity and position angle are derived using the {\sc idl} routine {\it find\_galaxy.pro} \citep{Krajnovic2011}. The ellipticity and position angle at a given radius are determined from the second moments of the luminosity distribution of all connected pixels above a given flux level. The values of R$_e$, $\epsilon$ and PA$\rm{_{phot}}$ for each galaxy are given in Table \ref{tab:properties}. The MGE fitting procedure failed for one spiral galaxy in the sample, J215705.29-071411.2, a LTG which is included in Table \ref{tab:properties} for completeness but excluded from all subsequent analysis.

\subsection{Kinematic Position Angles}
\label{sec:kin_pa_meas}

To measure the kinematic position angles (PA$\rm{_{kin}}$), and associated measurement errors, we use the {\it fit\_kinematic\_pa.pro} routine by Michele Cappellari, as outlined in Appendix C of \citet{Krajnovic2006}. The correct PA is found by comparing a bi-symmetric map based on the data to the data itself and minimising the residuals between the two, while varying the PA of the bi-symmetric map.

The velocity maps and kinematic PAs for the 79 ETGs in our sample are shown in Figure \ref{fig:mosaic_etg_pa}, with the kinematic PAs indicated in black and the photometric PAs in green. In the majority of cases the PAs are well aligned, though in some there is a clear discrepancy. This is discussed further, and quantified, in Section \ref{sec:kin_align}. The kinematic positions angles for the entire SAMI Pilot Survey sample are presented in Table \ref{tab:properties}.

\subsection{Calculating $\lambda_{R}$ and Kinematic Classification}
\label{sec:lr}

The kinematic classification of the SAMI Pilot Survey ETGs is discussed extensively in \citet{Fogarty2014}. We adopt the ATLAS$^{\rm{3D}}$ system using the $\lambda_{R}$ parameter, a proxy for the luminosity-weighted specific stellar angular momentum for each galaxy within a fiducial radius. $\lambda_{R}$ is defined as follows \citep{Emsellem2007}:

\begin{equation}
\lambda_R\equiv\frac{\langle R|V|\rangle}{\langle R\sqrt{V^2+\sigma^2}\rangle}=\frac{\Sigma_{i=0}^{N}F_i R_i |V_i|}{\Sigma_{i=0}^{N}F_i R_i \sqrt{V_i^2+\sigma_i^2}}
\label{eqn:lr}
\end{equation}

\noindent where $F_i, R_i, V_i, \sigma_i$ are the flux, radius, velocity and velocity dispersion of the $i$th of $N$ spaxels included in the sum. Since $\lambda_{R}$ is an integrated quantity it must be measured within a fiducial radius. For the SAMI Pilot Survey galaxies we use three fiducial radii, R$_e/2$, R$_e$, and 2R$_e$, chosen according to the size of the galaxy and how that compares to the SAMI field of view and the average seeing for the observations ($1.9\arcsec-3.0\arcsec$). For three of our galaxies $\lambda_{R}$ is measured within 2R$_e$ and for 15 $\lambda_{R}$ is measured within R$_e/2$. For the remainder $\lambda_{R}$ is measured within R$_e$. In \citet{Fogarty2014} we show that this does not impact on our FR/SR classifications. However, when considering the relative values of $\lambda_{R}$ for different kinematic classes the aperture effect in $\lambda_{R}$ must be addressed. This is discussed in detail in Section \ref{sec:dist_lam} where we compare the distribution of $\lambda_{R}$ between kinematic classes. 

For most galaxies in the sample we use the ellipticity and PA$\rm{_{phot}}$ measured at the fiducial radius to define the ellipse within which $\lambda_{R}$ is measured. However, we find seven galaxies with strong bars which compromise the measurements of the galaxy PA$\rm{_{phot}}$ and ellipticity at the fiducial radius. For those seven objects we use the radial profiles produced as described in Section \ref{sec:phot_pars} to instead measure the ellipticity and PA$\rm{_{phot}}$ of the galaxy in the outermost radial bins, away from the influence of the bar.

We classify the ETGs in our sample kinematically, as FRs and SRs, using the dividing line put forward by \citet{Emsellem2011}. This framework says that a galaxy is considered a SR if 

\begin{equation}
\lambda_{R}<k\sqrt{\epsilon}
\label{eqn:line}
\end{equation}

\noindent where $\lambda_{R}$ and $\epsilon$ are measured within the same fiducial radius and the proportionality constant $k$ is dependent on the choice of fiducial radius (see \citet{Fogarty2014} for details). We improved our method of fitting kinematics, accounting for possible template mismatch issues (see Section \ref{sec:stellarkin}). Therefore the values of $\lambda_{R}$ used here are very slightly different to those reported in \citet{Fogarty2014}. The average absolute difference is $\sim0.02$, within the average error on the quantity. This small difference is enough to alter the kinematic classification for two objects in the sample (J011421.54+001046.9 and J011515.78+004555.2). This does not alter any of the results in \citet{Fogarty2014}. Here we use the new values and new kinematic classifications, as reported in Table \ref{tab:properties}.

\begin{figure*}
\centering
\includegraphics[width=\textwidth]{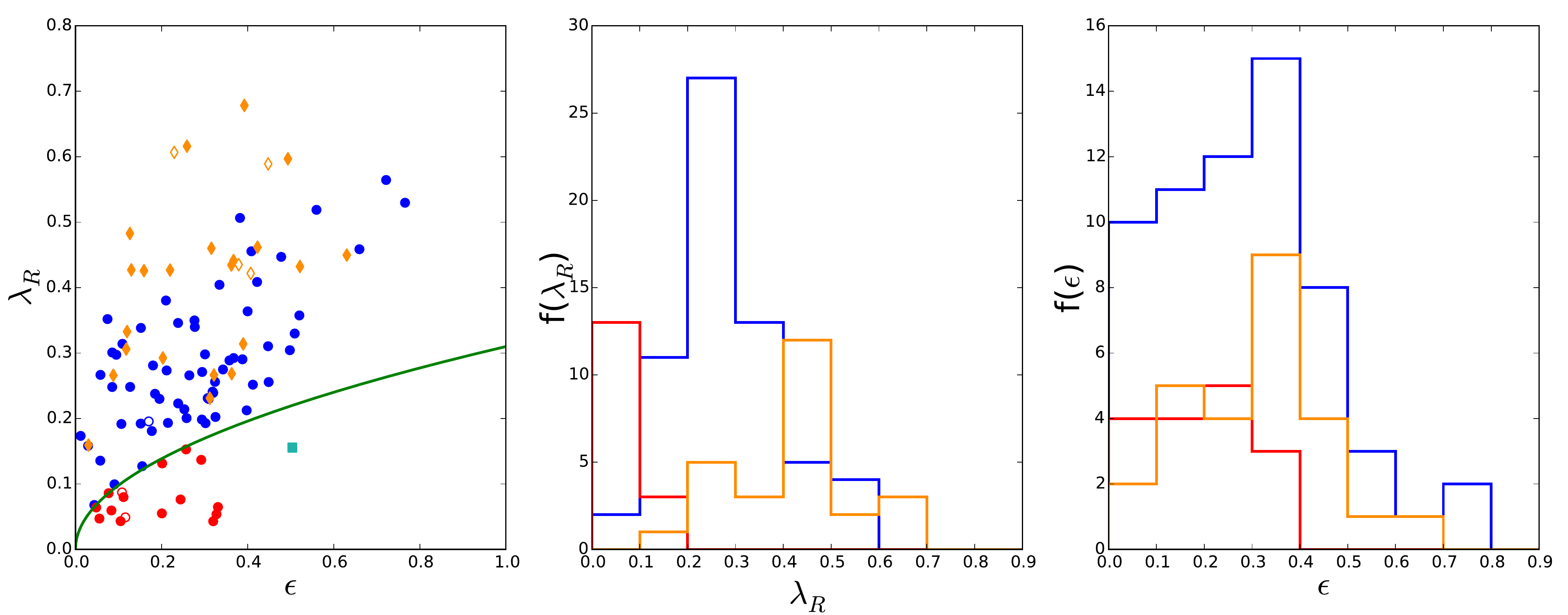}
\caption{The left panel shows the $\lambda_{R}-\epsilon$ plot for the SAMI Pilot Survey galaxies. The red and blue points are the ETGs, classified as SRs (red) or FRs (blue). The orange diamonds indicate the LTGs. The turquoise square indicates a single ``double sigma" galaxy, a contaminant to the SR population which is not a true SR. The filled symbols represent cluster members and the empty symbols represent non-members. The middle panel shows a histogram of $\lambda_{R}$ values for the SAMI Pilot Survey galaxies. The red outline denotes the SRs, the blue the FRs and the orange shows the LTGs. The peaks of the histograms for each class are clearly offset from one another, indicating a change in specific stellar angular momentum between the classes. The right panel shows a histogram of ellipticity values for the SAMI Pilot Survey galaxies, with colours as for the middle panel.}
\label{fig:lam_ltgs}
\end{figure*}

The $\lambda_{R}-\epsilon$ diagram is shown in the left-hand panel of Figure \ref{fig:lam_ltgs}. The ETGs are shown as circles and are colour coded according to their kinematic classification, with FRs in blue and SRs in red. Within the ETG sample we find 13 SRs.

For LTGs the classification of FR/SR is less relevant as LTGs are disks and are thus rotation supported systems with high stellar angular momentum. We find that all of the LTGs in our sample unsurprisingly meet the criterion to be classified as FRs (see the left-hand panel of Figure \ref{fig:lam_ltgs}). The $\lambda_{R}$ values for all of the SAMI Pilot Survey galaxies are given in Table \ref{tab:properties}.

To avoid confusion, for the remainder of this paper we discuss three different classes of galaxies. The LTGs, visually classified by their morphology; the FRs, first visually classified as ETGs and then further separated on the basis of the $\lambda_{R}-\epsilon$ diagram; and SRs, again first visually classified as ETG and then further separated kinematically. Thus the ETGs are made up of FRs and SRs and we do not consider any kinematic classification for the LTGs.

\section{Stellar Angular Momentum as a Function of Morphological Type}
\label{sec:ltgs}

Understanding the distribution of stellar angular momentum of galaxies with different morphology is crucial to understanding any evolutionary relationship between morphological types. ETGs are thought to form from LTGs through processes as diverse as mild harassment through to disruptive major mergers. How the angular momentum of a galaxy is modified by these processes can help us to determine which processes dominate at which times and in which environments.

We investigate these questions through the $\lambda_{R}$ proxy for projected specific stellar angular momentum. In this section we will discuss three classes of galaxies, the ETGs, kinematically separated into SRs and FRs, and the LTGs. The latter category contains 27 out of 106 galaxies in the SAMI Pilot sample for which kinematics are measured. Since this is a small number of galaxies no further subdivision into the usual Sa, Sb, Sc and Sd categories is attempted. 

\subsection{The Distribution of Stellar Angular Momentum}
\label{sec:dist_lam}

A histogram of $\lambda_{R}$ values for the SRs, FRs and LTGs in the SAMI Pilot Survey sample is shown in the middle panel of Figure \ref{fig:lam_ltgs}. The histogram peaks are clearly offset from one another, with median values of $\lambda_{R}$ of 0.27 and 0.43 for the FRs and LTGs respectively (we do not discuss the SRs as they are classified using $\lambda_{R}$). A K--S test rejects the hypothesis that the distribution of LTGs and FRs are drawn from the same parent sample, with a p-value of $2\times10^{-5}$. However, the fraction of LTGs where $\lambda_{R}$ is measured at R$_e/2$ is 35\%, whereas this fraction is only 0.5\% for the FRs, so this direct comparison is not quite fair. Since for both FRs and LTGs the radial $\lambda_{R}$ profiles tend to rise quickly in the inner parts comparing our sample of LTGs to a matched sample of FRs where 35\% of $\lambda_{R}$ values were measured at R$_e/2$ is likely to enhance to difference between the two distributions. To test this we measured $\lambda_{R}$ at R$_e/2$ for all FRs for which this was feasible and then randomly generated a sample of 63 individual FRs, wherein 35\% of the measurements were performed at R$_e/2$. This was repeated 100 times and the resulting average K--S p-value is $4.3\times10^{-6}$. This implies that there is a systematic difference in stellar angular momentum between the LTGs and the ETGs. 

However, $\lambda_{R}$ is a projected quantity and so it is possible that the difference in the distribution of $\lambda_{R}$ between the LTGs and FRs is simply driven by an underlying difference in the ellipticity distributions of the samples, reflecting an underlying difference in the inclination of the samples. The ellipticity distributions for the LTGs, FRs and SRs are shown in the right-hand panel of Figure \ref{fig:lam_ltgs}. The ellipticity distributions of the LTGs and FRs do not match each other closely, though they cannot be ruled out as being the same, with a K--S test yielding a p-value of 0.60. This suggests that although some of the difference in the $\lambda_{R}$ distributions of the two samples could be driven by projection effects, it is likely that the LTGs are a class of galaxies with higher specific stellar angular momentum than that of FRs.

Another effect that could drive this difference in $\lambda_{R}$ between the LTGs and FRs is a radial variation in mass to light ratio (M/L). This variation is potentially stronger for the LTGs, which tend to be brighter in the outskirts with lower M/L. This could be mitigated by calculating $\lambda_{R}$ using a mass weighting rather than a flux weighting. For this work we assume this effect is small and do not implement this correction.

Thus we find that LTGs have the higher angular momentum than the disk-supported ETGs, the FRs. This is expected if the evolution of LTGs to FRs occurs at least partly by dynamical processes such as tidal stripping, interactions or mergers, which can lower stellar angular momentum. 

The $\lambda_{R}-\epsilon$ diagram for the entire sample is shown in the left-hand panel of Figure \ref{fig:lam_ltgs}, where the cluster members are shown with filled symbols and the non-members with empty symbols. The difference between the classes of galaxies is less clear here and there is significant overlap between the LTGs and the FRs (seen also in the histograms in the middle and right-hand panels). The locus of points is slightly higher on the plane for the LTGs than the FRs.  

\subsection{Morphological Transformation}

\begin{figure}
\centering
\includegraphics[width=8.5cm]{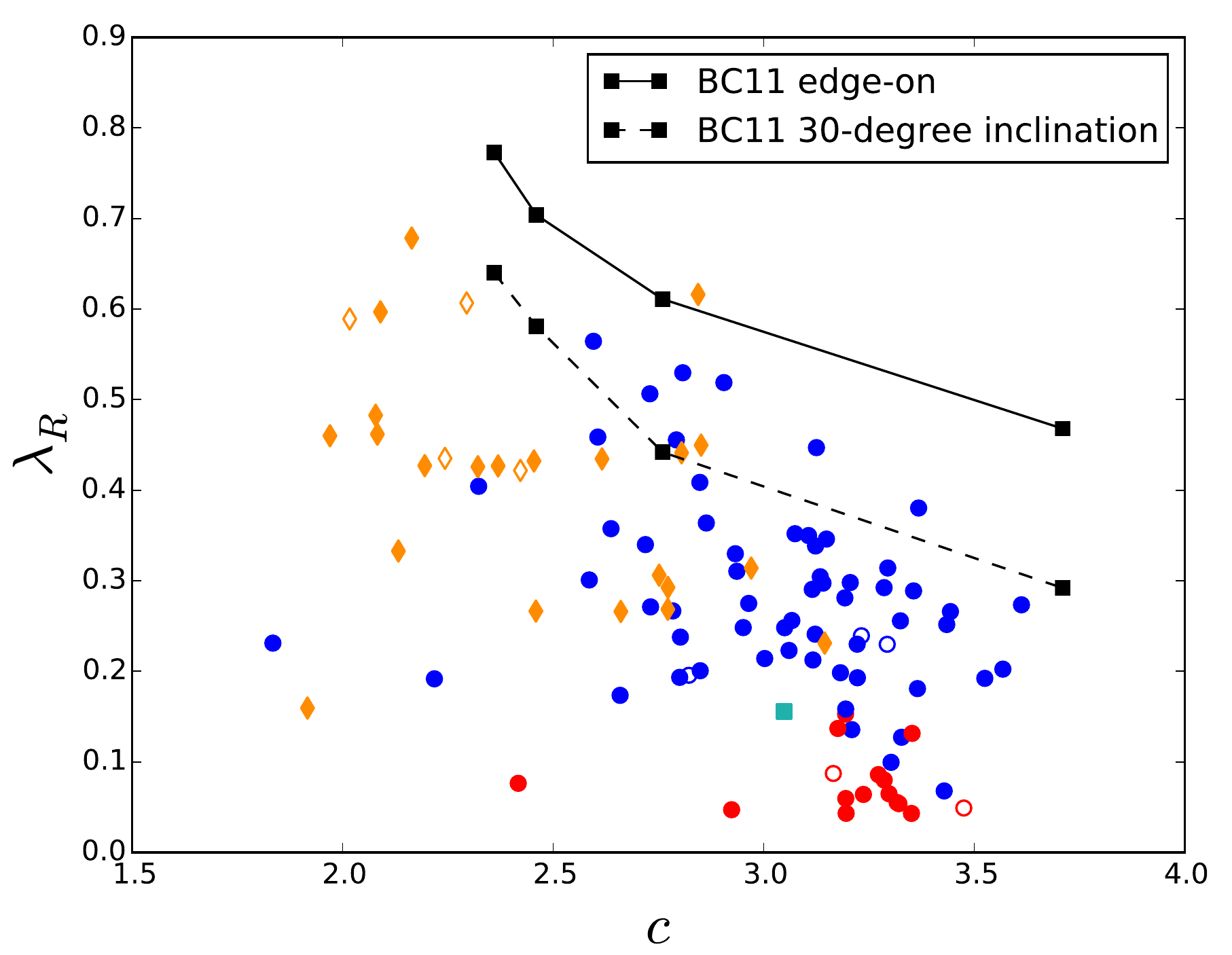}
\caption{$\lambda_{R}$ as a function of concentration for the SAMI Pilot Survey galaxies. The points are colour coded as in the left-hand panel of Figure \ref{fig:lam_ltgs}, with filled symbols representing cluster members and empty symbols non-members. The black track shows the evolutionary path of the fiducial model in \citet{BC11}, in which a spiral progenitor is subjected to multiple interactions with other galaxies and is thereby transformed into an S0.}
\label{fig:conc_ltgs}
\end{figure}

To investigate the distribution of angular momentum more deeply we examine $\lambda_{R}$ as a function of galaxy concentration, $c$, given by the ratio of the $r$-band Petrosian radius enclosing 90\% of the galaxy light to the $r$-band Petrosian radius enclosing 50\% of the galaxy light [$c=R_{90}/R_{50}$, from SDSS DR10; \citet{SDSSDR10_2014}]. The choice of photometric band does affect the resulting value for concentration, but we find that using the $g$, $r$, or $i$ bands do not remove the trend seen in the $\lambda_{R}-c$ plane and so we choose $r$-band for consistency with the rest of our anlysis. The results are shown in Figure \ref{fig:conc_ltgs}, where, again, cluster members are represented by closed symbols and non-members by open symbols. 

There is a trend from LTGs (orange) to FRs (blue) such that the FRs seem to have both lower specific angular momentum and higher concentration values. This is similar to that seen by \citet{Falcon-Barroso2015} using the CALIFA survey. FRs, being ETGs, will tend to have larger bulge-to-total luminosity ratios than the LTGs and thus higher concentration. The difference in specific angular momentum between the classes could be due to various dynamical processes involved in the formation of FRs from LTGs. The SRs on the other hand, have almost exclusively high concentration and, by definition, low angular momentum.

Figure \ref{fig:conc_ltgs} compares the observational SAMI results with the simulated evolution of a spiral galaxy transforming into an S0 galaxy (FR) \citep{BC11} on the $\lambda_{R}-c$ plane, for two different galaxy inclinations (the edge-on view, shown by the solid black track in Figure  \ref{fig:conc_ltgs} and a view with 30$^{\circ}$ inclination, shown by the the dashed black track). In both cases we plot the results for the fiducial MW-type model from \citet{BC11} with a total mass of $10^{12} {\rm M}_{\odot}$, moving in a group environment  with a total mass of $2 \times 10^{13} {\rm M}_{\odot}$. In this model the progenitor spiral has a bulge mass which is 17\% of the mass of the disk. The $c$ parameter of the simulated FR is derived by estimating a total stellar mass within a given radius (assuming a constant mass-to-light ratio). The spin parameter $\lambda_{R}$ for a simulated galaxy ($\lambda_{R, \rm{sim}}$) is derived from the simulated stellar kinematics shown in Figure 6 of \citet{BC11} as follows:

\begin{equation}
\lambda_{R,\rm{sim}}=\frac { \sum_{i=0}^{N} R_{i}|V_{i}| }
{ \sum_{i=0}^{N} R_{i}\sqrt{V_i^2+\sigma_i^2}},
\end{equation}

\noindent where $R_i$, $V_i$, $\sigma_i$ are the radius, velocity, and velocity dispersion, respectively, for each radial bin. This differs from the calculation of $\lambda_{R}$ from observed data as the luminosity factor ($F_i$ in Equation \ref{eqn:lr}) used in the observational $\lambda_{R}$ calculation is not included. The stars within $0.2R_{\rm d}$, where $R_{\rm d}$ is the initial disk size (=\,17.5 kpc), are used to derive $\lambda_{\rm R, sim}$ for the simulated galaxy. The value of $0.2R_{\rm d}$ is similar to the half-mass radius of the galaxy.

The edge-on model track from \citet{BC11} (solid line in Figure \ref{fig:conc_ltgs}) indicates that the progenitor spiral begins with high $\lambda_{R}$ (0.773) and a concentration of $c=2.36$, with the remnant FR having $\lambda_{R}=0.468$ and $c=3.71$. We have no galaxies in our sample with $\lambda_{R}$ as high as the edge-on model spiral, though we do not have any perfectly edge-on systems with which to compare. The dashed line in Figure \ref{fig:conc_ltgs} shows the track for the model with an inclination of 30 degrees. This galaxy starts with $\lambda_{R}=0.640$ and the remnant FR has a value of $\lambda_{R}=0.292$. This track matches more closely with our measured points. The initial concentration for the model spiral is close to the median value of 2.42 for our LTG sample (as opposed to a median of 3.12 for the FR sample). 

As the model evolves the simulated galaxy loses angular momentum due to repeated interactions with group members. This dynamically heats the galaxy disk causing the line-of-sight velocity to drop and the velocity dispersion to increase, lowering $\lambda_{R}$. By the same mechanism the bulge is grown via multiple episodes of star formation which coincide with galaxy-galaxy interactions \citep{BC11}, lowering the concentration, $c$.

Neither of the model tracks match our data exactly, however both exhibit trends in the same direction as the trend between the observed LTG and FR populations. This implies that the formation of some observed FRs could proceed via multiple encounters with other galaxies, as modelled in \citet{BC11}. However, there is a clear overlap between the LTG and FR populations in the $\lambda_{R}-c$ plane which may be caused by several complicating factors. 

Some FRs may not form through harassment (which would lower their angular momentum) but may instead maintain high $\lambda_{R}$ throughout their history. A process that could cause this is ram-pressure stripping, which removes gas from a galaxy causing star formation to cease, but does not significantly alter the galaxy dynamics. Many diverse evolutionary links between LTGs and FRs can be imagined and it will take detailed comparison to galaxy formation models to disentangle different effects. Such work is beyond the scope of this paper, but this preliminary result with a small sample of galaxies is promising and indicates that dynamical effects are likely to be important for the formation of at least some of the FR population. 

\section{Kinematic Misalignment}
\label{sec:kin_align}

LTGs and FRs are classes of disk galaxies with many similar properties. SRs, on the other hand, are thought to be a separate class of galaxies with very different intrinsic properties. \citet{Krajnovic2011} and \citet{Weijmans2014} find that SRs are not oblate spheroids, like disk galaxies, but show evidence of mildly triaxial figures. This implies that  the formation histories of FRs and SRs will likely be different.

We compare the photometric and kinematic PAs ($\rm{PA_{phot}}$ and $\rm{PA_{kin}}$) of the ETGs in the SAMI Pilot sample. A misalignment between these two PAs can be indicative of the intrinsic shape of a galaxy, with oblate spheroids (disks) usually displaying a close alignment between these two quantities. Triaxiality, on the other hand, can introduce a misalignment.

\begin{figure}
\centering
\includegraphics[width=8.5cm]{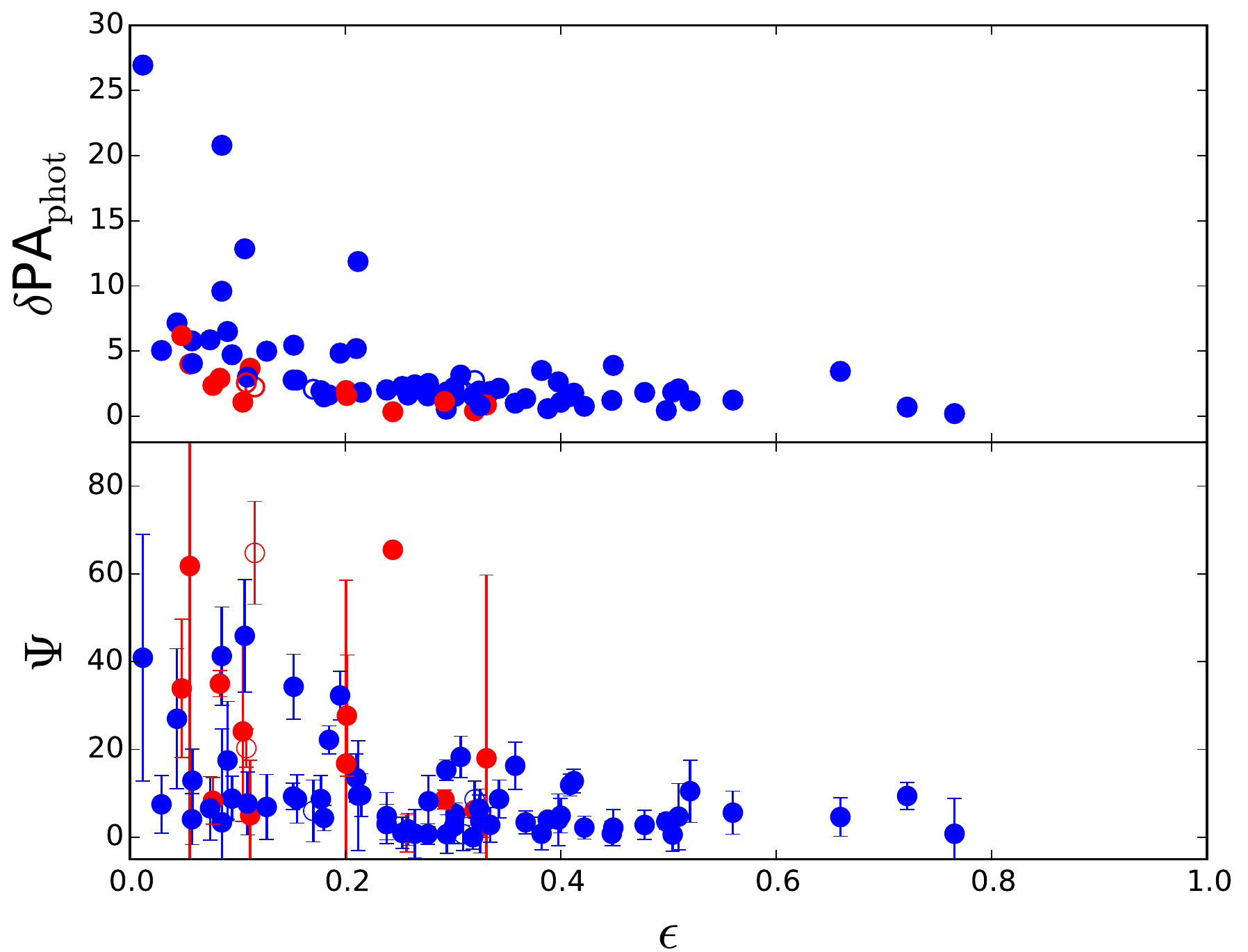}
\caption{The top panel shows the uncertainty in the measured photometric PA ($\delta \rm{PA_{phot}}$) as a function of ellipticity. There is a systematic trend towards larger uncertainty for rounder galaxies. The bottom panel shows the kinematic misalignment angle $\Psi$ as a function of ellipticity. The FRs are in blue and the SRs in red, with the cluster members indicated by filled symbols and the non-members by empty symbols.}
\label{fig:pa_phot_err}
\end{figure}

The values of $\rm{PA_{phot}}$ and $\rm{PA_{kin}}$ are calculated as described in Section \ref{sec:der_params}. The measurement uncertainties on these quantites are also derived. The mean uncertainty in the photometric PA is 4.1$^\circ$ and the median uncertainty is 2.2$^\circ$, though there is a tail in the distribution towards large uncertainty. This is a systematic effect as it is difficult to constrain PA$\rm{_{phot}}$ for objects with lower ellipticity. This is illustrated in the top panel of Figure \ref{fig:pa_phot_err}.

\begin{figure*}
\centering
\includegraphics[width=17cm]{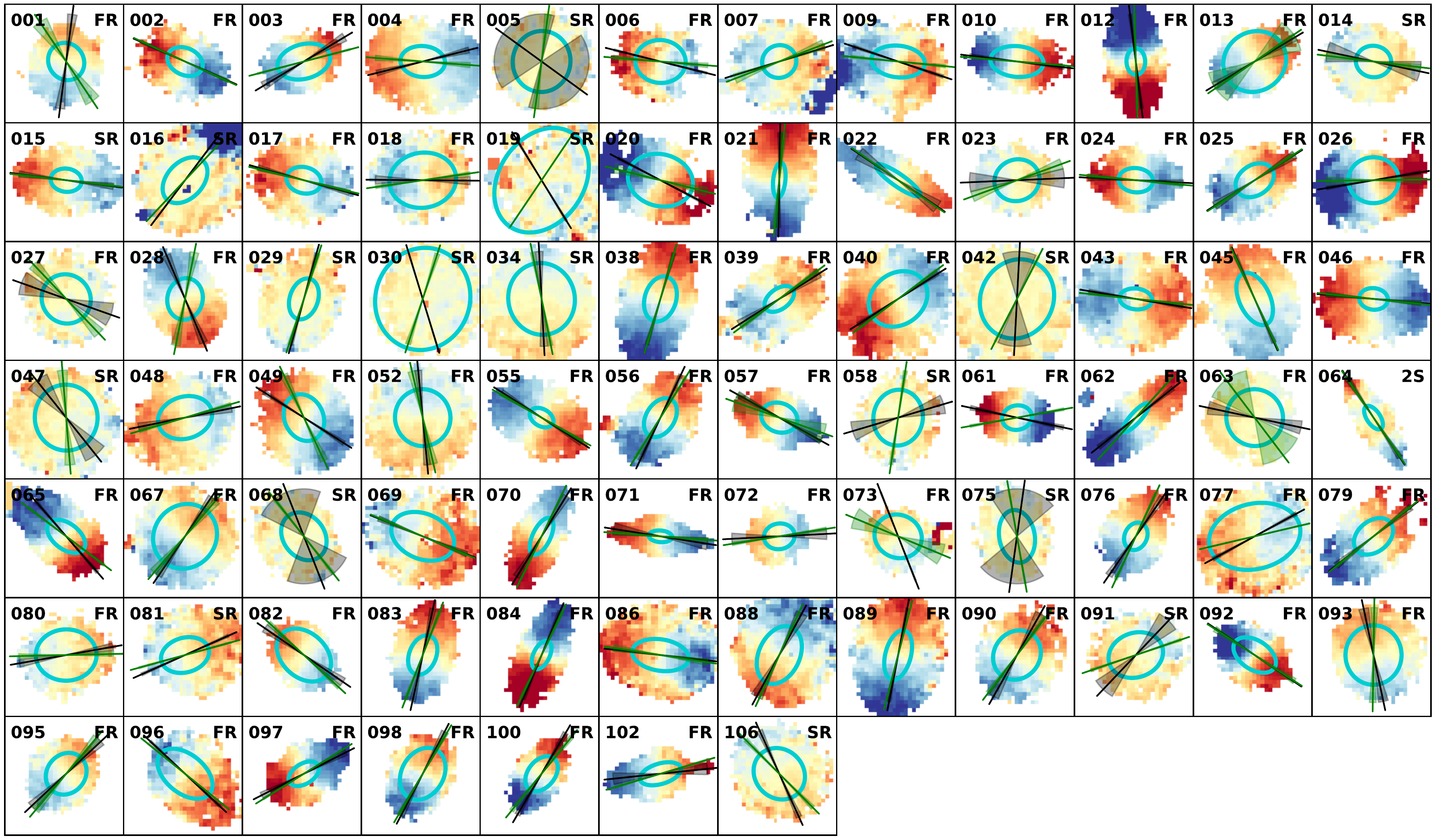}
\caption{The stellar velocity fields are shown for the 79 ETGs in the SAMI Pilot Survey. For comparison each panel is shown with the same scale, from -200/+200 km\,s$^{-1}$ and in each panel North is up and East is to the left. The black line and shaded black wedge represent the best fit kinematic PA, \pakin, and the measurement error on \pakin\, respectively. The green line and shaded green wedge represent the best fit photometric PA, \paphot. In each panel the turquoise ellipse represents the fiducial radius within which $\lambda\rm{_R}$ was measured. The fiducial radius chosen for each galaxy is shown in Table \ref{tab:properties}.}
\label{fig:mosaic_etg_pa}
\end{figure*}

Figure \ref{fig:mosaic_etg_pa} shows the kinematic maps for the 79 ETGs, overlaid with thei]r kinematic (black) and photometric PAs (green). The mean uncertainty in the kinematic PA is 6.3$^\circ$ and the median is 3.75$^\circ$, though as can be seen from Figure \ref{fig:mosaic_etg_pa} there is a large tail in this distribution towards large uncertainty, similar to that seen in the distribution of errors on \paphot. This is also a systematic effect as objects with low or no rotation do not have a well-constrained kinematic PA. 

Following \citet{Krajnovic2011} and \citet{Franx1991} we calculate the kinematic misalignment angle $\Psi$ for the 79 ETGs in our sample. $\Psi$ is defined as the difference between \paphot\, and \pakin\,:

\begin{equation}
\rm{sin}\Psi=|\rm{sin}(\rm{PA_{phot}-PA_{kin}})|.
\end{equation}

\begin{figure}
\centering
\includegraphics[width=8.5cm]{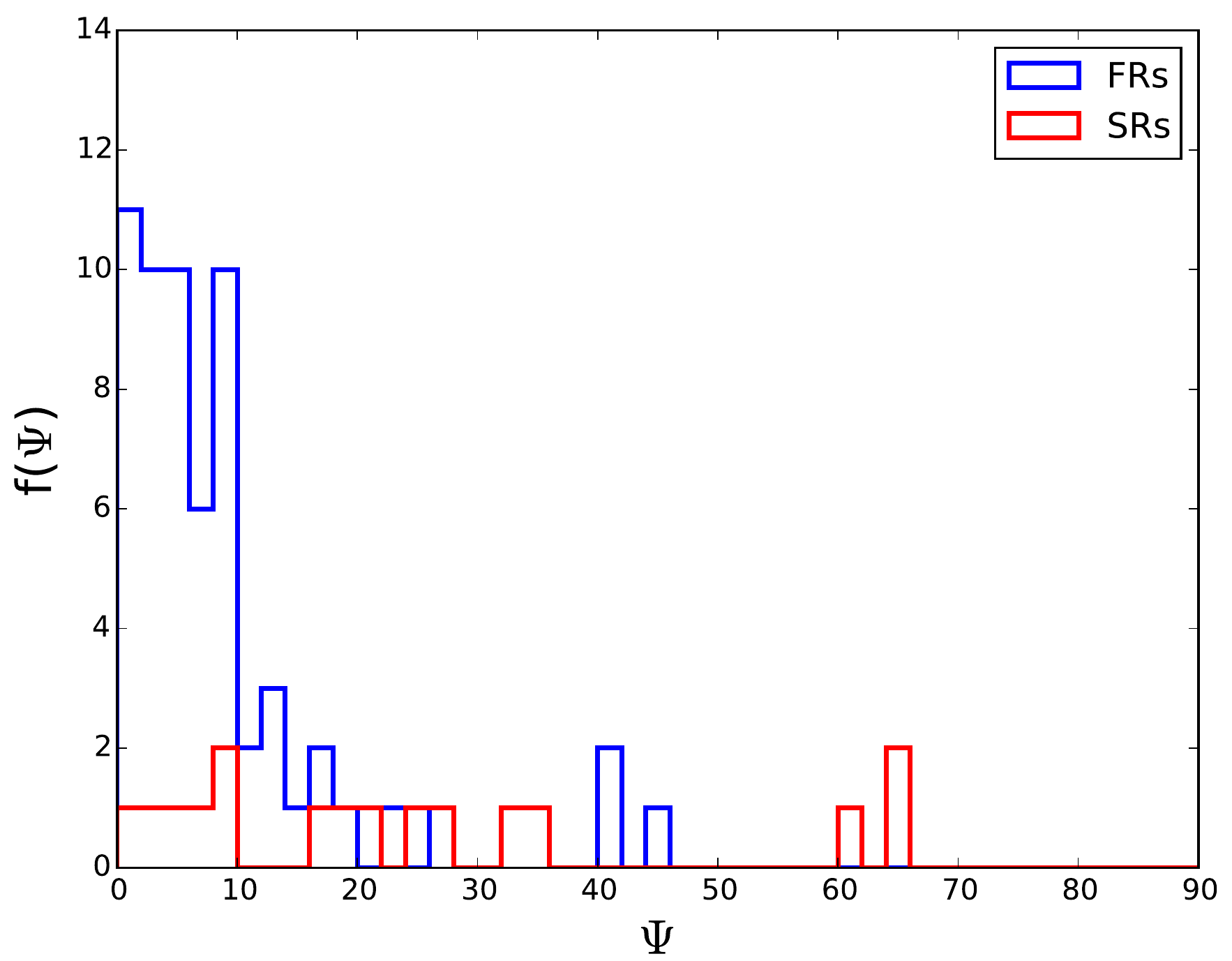}
\caption{A histogram of the magnitude of kinematic misalignment, $\Psi$, for the ETGs in the SAMI Pilot sample. The blue histogram indicates the FRs in the sample and the red indicates the SRs.}
\label{fig:frsr_pa}
\end{figure}

Using this definition, $\Psi$ lies between 0$^\circ$ and 90$^\circ$ and is insensitive to differences of 180$^\circ$ between the two PAs. The value of $\Psi$ for each of the ETGs is given in Table \ref{tab:properties}, and $\Psi$ is shown as a function of ellipticity in the bottom panel of Figure \ref{fig:pa_phot_err}. A histogram is shown in Figure \ref{fig:frsr_pa}, with the ETGs split kinematically. FRs are shown in blue and SRs in red. A K--S test indicates that the distributions in $\Psi$ are not the same for FRs and SRs with a p-value of $3.2\times10^{-3}$ (if we restrict our analysis to include only those FRs with $\epsilon\leq0.4$, matching the ellipticity distribution of the SRs, the result stands with a K--S test p-value of $1.2\times10^{-2}$). This implies that the two classes of galaxies have different distributions in alignment and therefore in intrinsic shape.  

In their sample of 260 galaxies from the ATLAS$^{\rm{3D}}$ survey, \citet{Krajnovic2011} found that 90\% of galaxies were aligned, with $\Psi< 15^\circ$. If we adopt the ATLAS$^{\rm{3D}}$ constraint we find that 78\% of ETGs have $\Psi< 15^\circ$. However, some galaxies have large error bars on the measured position angles and so we also count as aligned any galaxies where the \paphot\, and \pakin\, error bars overlap. This gives us an aligned fraction of 73\%. If we split our sample by kinematic class then we find that within the sample of FRs 83\% are aligned systems. However, within the SR population the aligned fraction is only 38\%. 

These result gives a strong indication that SRs have an intrinsically different shape distribution than FRs.

\section{Discussion}
\label{sec:discussion}

\subsection{The Distribution of Angular Momentum}

The results presented in Section \ref{sec:ltgs} show that the average measured proxy for projected specific stellar angular momentum, $\lambda_{R}$, is different for FRs and LTGs. 

Much work has been done to investigate an evolutionary link between LTGs and FRs (e.g. \citet{GunnGott1972}, \citet{Larson1980} and \citet{Bekki1998}). Such an evolutionary link could be one reason for the trend we see from LTGs to FRs in the $\lambda_{R}-c$ plane.  

The overlap between the LTG and FR angular momentum distributions could indicate that there are many mechanisms by which FRs can be formed from LTGs. Different evolutionary paths could affect the position of the remnant galaxy in the $\lambda_{R}-c$ plane differently.

The example from \citet{BC11} shown in Figure \ref{fig:conc_ltgs} shows one type of dynamical interaction, namely repeated interactions within a galaxy group, which grows a bulge and changes the angular momentum of a galaxy. Major and minor mergers could have a similar effect, growing a bulge and lowering the angular momentum of the remnant FR \citep{Khochfar2011, Naab2013}.

Morphological transformation can also occur by secular processes within galaxies. For example, passive fading of LTGs to FRs, by the exhaustion of gas and cessation of SF, will likely create FR remnants with almost unchanged angular momentum, though as $\lambda_{R}$ is a flux-weighted quantity it will not be completely unchanged. Passive fading could cause a change in $c$ due to the blue star-forming disk fading, with the underlying bulge component coming to dominate the light. These effects will likely be smaller than those from a merger or other interaction, resulting in little movement in the $\lambda_{R}-c$ plane as a result.

Disk instabilities can grow a galactic bulge, though a bulge generated in this way, a secular bulge, will likely have some different properties to those generated through dynamical interactions \citep{Kormendy2004}. It is not clear how the processes which create a secular bugle would change the angular momentum of the resulting galaxy. Secularly grown bulges tend to be more disk-like in nature, and have been shown by \citet{Fabricius2012} to have higher rotational support than classical bulges. Thus a galaxy with a secular bulge could end up with higher $\lambda_{R}$ for a given concentration than a galaxy with a classical bulge. This may be one way to achieve a large spread in $\lambda_{R}$ within LTGs and FRs.

The processes described above illustrate a variety of different evolutionary paths between LTGs and FRs which could account for the observed trend in the $\lambda_{R}-c$ plane. It is worth noting that the same trend could be seen if galaxies with differing bulge-to-total mass ratios simply have different angular momentum content -- with no need for an evolutionary link between the different classes of galaxies. However, the similarity between the trends seen in the model from \citep{BC11} and our data indicates morphological transformation through dynamical processes are likely to be important when explaining the distribution of galaxies in the $\lambda_{R}-c$ plane.

Lastly, different FR formation processes are thought to dominate in different environments. Since our sample consists mostly of cluster member galaxies we cannot explore this further. However, the SAMI Galaxy Survey will have $\sim3400$ galaxies, about a quarter of which lie in clusters, with the remainder occupying smaller groups and the field. With a sample this large it will be possible to construct the $\lambda_{R}-c$ plane for a range of environments and stellar masses. It may then be possible to isolate the importance of different processes in various environments.

\subsection{Galaxy Shapes}

In Section \ref{sec:kin_align} we showed that the two kinematic classes of ETGs, the FRs and SRs, have different distributions in misalignment angle, $\Psi$. The FRs are rotation supported systems with high stellar angular momentum and the majority show close alignment between the kinematic and photometric PAs. This is expected if these galaxies are oblate spheroids.

The SRs on the other hand are pressure-supported systems with low stellar angular momentum. They display a wider range of $\Psi$ values and only 38\% are aligned systems. This implies that these galaxies are not oblate spheroids. An intrinsically triaxial shape for these galaxies could cause this. 

Our current sample of ETGs is small, so it is difficult to further constrain this problem. This is true especially in the case of SRs, of which we observe only thirteen. \citet{Weijmans2014} perform an inversion of the observed ellipticity distributions for their samples of FRs and SRs from the ATLAS$\rm{^{3D}}$ survey to probe the intrinsic shape distribution for each class of galaxy. This kind of analysis is not possible with the SAMI Pilot Survey. However as the full SAMI Galaxy Survey will comprise $\sim3400$ galaxies, it is expected that roughly 200 of those likely to be SRs, increasing our current sample by more than an order of magnitude. This will allow a full analysis of the shape distribution of ETGs for a large sample of galaxies.

\section{Conclusions}
\label{sec:conclusions}
We present the SAMI Pilot Survey, an IFS survey of 106 galaxies in three galaxy clusters. We have derived stellar velocity and velocity dispersion maps for all 106 galaxies and these are presented in Appendix B. 

We also present SAMI observations of two test galaxies drawn from the ATLAS$^{\rm{3D}}$ survey (NGC1289 and NGC1665). Although the observation procedure was specific to these galaxies the SAMI data reduction procedure could be used with only minor changes. The data were analysed to extract stellar kinematics in the same way as for the SAMI Pilot Survey. The SAMI velocity and velocity dispersion maps of NGC1289 and NGC1665 match those from ATLAS$^{\rm{3D}}$ extremely well. We conclude that the SAMI data reduction procedure is robust, producing uniform data cubes capable of reproducing known results.

We examine the distribution of angular momentum in our sample using $\lambda_{R}$, a proxy for the projected specific stellar angular momentum  in a galaxy. We find that the median $\lambda_{R}$ for LTGs is higher than that for FRs, although there is significant overlap between the two populations. We find a trend in the $\lambda_{R}-c$ plane such that LTGs have higher $\lambda_{R}$ and lower concentrations. We suggest that this could be due to an evolutionary link between these groups. The observed trend is consistent with a model wherein a LTG is transformed into a FR through repeated interactions in a group environment. There is a large overlap between the LTG and FR distributions indicating that there are many processes which can affect where a galaxy lies in the $\lambda_{R}-c$ plane.

We find that a large percentage of the FRs in our sample (83\%) show kinematics which are aligned with their photometric PA. This is consistent with their interpretation as oblate spheroids. On the other hand only 38\% of the SRs are aligned. This implies that FRs are consistent with being oblate spheroids and SRs are consistent with mild triaxiality.

\section*{ACKNOWLEDGEMENTS} 

The SAMI Galaxy Survey is based on observations made at the Anglo-Australian Telescope. The Sydney-AAO Multi-object Integral field spectrograph (SAMI) was developed jointly by the University of Sydney and the Australian Astronomical Observatory. The SAMI input catalogue is based on data taken from the Sloan Digital Sky Survey, the GAMA Survey and the VST ATLAS Survey. The SAMI Galaxy Survey is funded by the Australian Research Council Centre of Excellence for All-sky Astrophysics (CAASTRO), through project number CE110001020, and other participating institutions. The SAMI Galaxy Survey website is http://sami-survey.org.

MSO acknowledges the funding support from the Australian Research Council through a Super Science Fellowship (ARC FS110200023) and through a Future Fellowship (FT140100255).

SMC acknowledges the support of an ARC future fellowship (FT100100457).

JBH acknowledges the support of an Australian Laureate Fellowship.

RCWH was supported by the Science and Technology Facilities Council [STFC grant numbers ST/H002456/1, ST/K00106X/1 \& ST/J002216/1].

LC acknowledges support under the Australian Research CouncilÕs Discovery Projects funding scheme 
(project number 130100664).

This work was supported by the Astrophysics at Oxford grants (ST/H002456/1 and ST/K00106X/1) as well as visitors grant (ST/H504862/1) from the UK Science and Technology Facilities Council. RLD acknowledges travel and computer grants from Christ Church, Oxford. RLD is also grateful for support from the Australian Astronomical Observatory Distinguished Visitors programme, the ARC Centre of Excellence for All Sky Astrophysics, and the University of Sydney during a sabbatical visit.

JTA and ISK are the recipients of John Stocker Postdoctoral Fellowships from the Science and Industry Endowment Fund (Australia).

This research made use of Montage, funded by the National Aeronautics and Space Administration's Earth Science Technology Office, Computational Technologies Project, under Cooperative Agreement Number NCC5-626 between NASA and the California Institute of Technology. The code is maintained by the NASA/IPAC Infrared Science Archive.

This research made use of Astropy, a community-developed core Python package for Astronomy (Astropy Collaboration, 2013)

\bibliography{SAMI_StellarKin}
\bibliographystyle{mn2e}

\appendix
\section[]{The SAMI Pilot Survey Galaxies}

\begin{table*}
\centering
\begin{tabular}{|c|c|c|c|c|c|c|c|c|c|c|c|c|}
\hline
Galaxy &Galaxy Name & Cluster & m & M$_r$ & $\epsilon$ & $\rm{PA_{phot}}$ & $\rm{PA_{kin}}$ & $\Psi$ & $\lambda_R$ & $\Delta\lambda_R$ & Galaxy & Fiducial \\
ID & & & & & & & & & & & Class & Radius\\\hline
001 & J003906.77-084758.3 & ABELL0085 & 1 & -20.66 & 0.085 & 33.8 & 172.5 & 41.3 & 0.248 & 0.015 & FR & Re\\
002 & J004001.68-095252.5 & ABELL0085 & 1 & -21.02 & 0.276 & 65.3 & 66.0 & 0.7 & 0.35 & 0.009 & FR & Re\\
003 & J004004.88-090302.6 & ABELL0085 & 1 & -20.72 & 0.358 & 104.7 & 121.0 & 16.3 & 0.289 & 0.011 & FR & Re\\
004 & J004018.68-085257.1* & ABELL0085 & 1 & -21.45 & 0.307 & 85.7 & 104.0 & 18.3 & 0.231 & 0.013 & FR & Re\\
005 & J004046.47-085005.0 & ABELL0085 & 1 & -21.04 & 0.056 & 172.2 & 54.0 & 61.8 & 0.047 & 0.009 & SR & Re\\
006 & J004101.87-091233.1 & ABELL0085 & 1 & -20.78 & 0.152 & 85.3 & 76.0 & 9.3 & 0.192 & 0.012 & FR & Re\\
007 & J004112.21-091010.2 & ABELL0085 & 1 & -21.02 & 0.058 & 111.1 & 107.0 & 4.1 & 0.136 & 0.007 & FR & Re\\
008 & J004112.79-093203.7 & ABELL0085 & 1 & -20.67 & 0.392 & 102.7 & 107.0 & 4.3 & 0.678 & 0.064 & LTG & Re\\
009 & J004122.06-095240.8 & ABELL0085 & 1 & -21.33 & 0.412 & 84.3 & 71.5 & 12.8 & 0.252 & 0.01 & FR & Re2\\
010 & J004128.56-093426.7 & ABELL0085 & 1 & -20.71 & 0.422 & 85.2 & 83.0 & 2.2 & 0.409 & 0.011 & FR & Re\\
011 & J004130.29-091545.8* & ABELL0085 & 1 & -21.39 & 0.362 & 3.6 & 5.0 & 1.4 & 0.435 & 0.023 & LTG & Re\\
012 & J004130.42-091406.7 & ABELL0085 & 1 & -21.04 & 0.302 & 1.7 & 7.0 & 5.3 & 0.193 & 0.012 & FR & Re\\
013 & J004131.25-094151.0 & ABELL0085 & 1 & -20.5 & 0.086 & 124.4 & 121.0 & 3.4 & 0.301 & 0.014 & FR & Re\\
014 & J004133.41-090923.4 & ABELL0085 & 1 & -20.8 & 0.112 & 83.0 & 78.0 & 5.0 & 0.08 & 0.01 & SR & Re\\
015 & J004134.89-092150.5 & ABELL0085 & 1 & -21.14 & 0.257 & 83.4 & 82.5 & 0.9 & 0.153 & 0.01 & SR & Re\\
016 & J004143.00-092621.9 & ABELL0085 & 1 & -21.62 & 0.32 & 136.8 & 143.0 & 6.2 & 0.043 & 0.01 & SR & Re2\\
017 & J004148.22-091703.1 & ABELL0085 & 1 & -20.86 & 0.258 & 76.3 & 74.5 & 1.8 & 0.201 & 0.009 & FR & Re\\
018 & J004150.17-092547.4 & ABELL0085 & 1 & -21.56 & 0.177 & 98.2 & 89.5 & 8.7 & 0.181 & 0.008 & FR & Re\\
019 & J004150.46-091811.2 & ABELL0085 & 1 & -22.92 & 0.244 & 146.0 & 31.5 & 65.5 & 0.076 & 0.009 & SR & Re2\\
020 & J004152.16-093014.8 & ABELL0085 & 1 & -21.36 & 0.21 & 76.0 & 62.5 & 13.5 & 0.38 & 0.008 & FR & Re\\
021 & J004153.50-092943.9 & ABELL0085 & 1 & -20.9 & 0.66 & 174.4 & 179.0 & 4.6 & 0.459 & 0.013 & FR & Re2\\
022 & J004200.64-095004.0 & ABELL0085 & 1 & -20.67 & 0.766 & 55.2 & 56.0 & 0.8 & 0.53 & 0.019 & FR & Re\\
023 & J004205.86-090240.7 & ABELL0085 & 1 & -20.67 & 0.091 & 110.0 & 92.5 & 17.5 & 0.1 & 0.011 & FR & Re\\
024 & J004215.91-093252.0 & ABELL0085 & 1 & -20.78 & 0.301 & 84.5 & 87.0 & 2.5 & 0.298 & 0.01 & FR & Re\\
025 & J004218.75-091528.4 & ABELL0085 & 1 & -20.6 & 0.265 & 123.3 & 122.5 & 0.8 & 0.266 & 0.011 & FR & Re\\
026 & J004233.86-091040.5 & ABELL0085 & 1 & -21.01 & 0.095 & 90.7 & 99.5 & 8.8 & 0.297 & 0.009 & FR & Re\\
027 & J004233.99-095442.2 & ABELL0085 & 1 & -21.24 & 0.044 & 43.5 & 70.5 & 27.0 & 0.068 & 0.01 & FR & Re\\
028 & J004242.26-085528.1 & ABELL0085 & 1 & -20.62 & 0.152 & 168.7 & 23.0 & 34.3 & 0.338 & 0.012 & FR & Re\\
029 & J004244.68-093316.3 & ABELL0085 & 1 & -21.0 & 0.328 & 161.9 & 164.5 & 2.6 & 0.054 & 0.007 & SR & Re\\
030 & J004310.12-095141.2 & ABELL0085 & 1 & -21.74 & 0.084 & 162.0 & 17.0 & 35.0 & 0.06 & 0.007 & SR & Re\\
031 & J011327.21+000908.9 & ABELL0168 & 1 & -20.96 & 0.12 & 166.7 & 152.5 & 14.2 & 0.333 & 0.015 & LTG & Re2\\
032 & J011346.32+001820.6* & ABELL0168 & 1 & -21.09 & 0.522 & 85.1 & 97.0 & 11.9 & 0.432 & 0.013 & LTG & Re2\\
033 & J011415.78+004555.2* & ABELL0168 & 1 & -20.97 & 0.03 & 52.1 & 53.0 & 0.9 & 0.159 & 0.025 & LTG & Re\\
034 & J011421.54+001046.9 & ABELL0168 & 1 & -21.12 & 0.077 & 11.3 & 3.0 & 8.3 & 0.086 & 0.006 & SR & Re\\
035 & J011425.68+003209.9 & ABELL0168 & 1 & -20.61 & 0.316 & 149.0 & 10.0 & 41.0 & 0.46 & 0.102 & LTG & Re2\\
036 & J011430.80+001928.3* & ABELL0168 & 1 & -21.1 & 0.088 & 49.8 & 21.5 & 28.3 & 0.266 & 0.012 & LTG & Re2\\
037 & J011443.86+001709.6 & ABELL0168 & 1 & -20.44 & 0.423 & 111.8 & 125.5 & 13.7 & 0.462 & 0.051 & LTG & Re2\\
038 & J011446.94+003128.8 & ABELL0168 & 1 & -20.68 & 0.319 & 163.5 & 163.5 & 0.0 & 0.241 & 0.006 & FR & Re\\
039 & J011454.21+003026.5 & ABELL0168 & 1 & -20.43 & 0.398 & 126.5 & 122.5 & 4.0 & 0.213 & 0.011 & FR & Re\\
040 & J011454.25+001811.8 & ABELL0168 & 1 & -21.01 & 0.18 & 126.9 & 122.5 & 4.4 & 0.281 & 0.006 & FR & Re\\
041 & J011456.26+000750.4 & ABELL0168 & 1 & -20.79 & 0.367 & 154.6 & 108.5 & 46.1 & 0.441 & 0.013 & LTG & Re\\
042 & J011457.59+002550.8 & ABELL0168 & 1 & -22.14 & 0.105 & 152.9 & 177.0 & 24.1 & 0.043 & 0.004 & SR & Re2\\
043 & J011459.61+001533.1 & ABELL0168 & 1 & -20.95 & 0.368 & 83.9 & 80.5 & 3.4 & 0.292 & 0.011 & FR & Re\\
044 & J011503.63+002418.7 & ABELL0168 & 1 & -21.2 & 0.363 & 109.0 & 110.0 & 1.0 & 0.269 & 0.009 & LTG & Re\\
045 & J011507.33+002756.8 & ABELL0168 & 1 & -20.55 & 0.448 & 24.8 & 24.0 & 0.8 & 0.31 & 0.009 & FR & Re\\
046 & J011508.73+003433.5 & ABELL0168 & 1 & -20.76 & 0.253 & 84.0 & 85.0 & 1.0 & 0.214 & 0.012 & FR & Re\\
047 & J011515.78+001248.4 & ABELL0168 & 1 & -21.15 & 0.048 & 4.6 & 38.5 & 33.9 & 0.064 & 0.007 & SR & Re\\
048 & J011516.77+001108.3 & ABELL0168 & 1 & -20.81 & 0.239 & 106.3 & 101.5 & 4.8 & 0.223 & 0.008 & FR & Re\\
049 & J011531.18+001757.2 & ABELL0168 & 1 & -20.92 & 0.195 & 25.2 & 57.5 & 32.3 & 0.23 & 0.007 & FR & Re\\
050 & J011603.31-000652.7* & ABELL0168 & 1 & -21.15 & 0.203 & 3.0 & 118.5 & 64.5 & 0.293 & 0.041 & LTG & Re\\
051 & J011605.60-000053.6 & ABELL0168 & 1 & -21.1 & 0.63 & 67.5 & 71.5 & 4.0 & 0.45 & 0.013 & LTG & Re2\\
052 & J011612.79-000628.3 & ABELL0168 & 1 & -20.98 & 0.029 & 13.0 & 5.5 & 7.5 & 0.158 & 0.008 & FR & Re\\
053 & J011623.61+002644.8 & ABELL0168 & 0 & -20.76 & 0.407 & 131.1 & 131.5 & 0.4 & 0.422 & 0.02 & LTG & Re\\
054 & J011703.58+000027.4 & ABELL0168 & 1 & -20.62 & 0.39 & 34.2 & 1.0 & 33.2 & 0.314 & 0.018 & LTG & Re\\
\hline
\end{tabular}
\end{table*}

\begin{table*}
\centering
\begin{tabular}{|c|c|c|c|c|c|c|c|c|c|c|c|c|}
\hline
Galaxy &Galaxy Name & Cluster & m & M$_r$ & $\epsilon$ & $\rm{PA_{phot}}$ & $\rm{PA_{kin}}$ & $\Psi$ & $\lambda_R$ & $\Delta\lambda_R$ & Galaxy & Fiducial \\
ID & & & & & & & & & & & Class & Radius\\\hline
055 & J215432.20-070924.1 & ABELL2399 & 1 & -20.5 & 0.335 & 60.3 & 57.5 & 2.8 & 0.404 & 0.036 & FR & Re2\\
056 & J215445.80-072029.2 & ABELL2399 & 1 & -20.94 & 0.324 & 148.0 & 154.5 & 6.5 & 0.256 & 0.009 & FR & Re\\
057 & J215447.94-074329.7 & ABELL2399 & 1 & -20.93 & 0.212 & 70.5 & 61.0 & 9.5 & 0.273 & 0.011 & FR & 2Re\\
058 & J215457.43-073551.3 & ABELL2399 & 0 & -21.52 & 0.116 & 171.3 & 106.5 & 64.8 & 0.049 & 0.009 & SR & Re\\
059 & J215556.95-065337.9 & ABELL2399 & 1 & -21.6 & 0.322 & 122.5 & 144.5 & 22.0 & 0.267 & 0.022 & LTG & Re\\
060 & J215604.08-071938.1 & ABELL2399 & 1 & -20.29 & 0.259 & 23.4 & 24.5 & 1.1 & 0.616 & 0.081 & LTG & Re\\
061 & J215619.00-075515.6 & ABELL2399 & 1 & -20.25 & 0.185 & 100.2 & 78.0 & 22.2 & 0.238 & 0.013 & FR & Re\\
062 & J215624.58-081159.8 & ABELL2399 & 1 & -20.85 & 0.722 & 137.9 & 128.5 & 9.4 & 0.564 & 0.017 & FR & Re\\
063 & J215628.95-074516.1 & ABELL2399 & 1 & -20.73 & 0.012 & 37.1 & 78.0 & 40.9 & 0.173 & 0.024 & FR & Re\\
064 & J215634.45-075217.5 & ABELL2399 & 1 & -20.29 & 0.504 & 34.1 & 33.5 & 0.6 & 0.156 & 0.019 & 2S & Re\\
065 & J215635.58-075616.9 & ABELL2399 & 1 & -20.69 & 0.409 & 52.9 & 41.0 & 11.9 & 0.456 & 0.011 & FR & Re\\
066 & J215636.04-065225.6 & ABELL2399 & 1 & -20.54 & 0.494 & 162.5 & 167.0 & 4.5 & 0.597 & 0.083 & LTG & Re2\\
067 & J215637.29-074043.0 & ABELL2399 & 1 & -22.41 & 0.074 & 139.4 & 146.0 & 6.6 & 0.352 & 0.011 & FR & Re\\
068 & J215643.13-073259.8 & ABELL2399 & 1 & -20.9 & 0.201 & 38.3 & 21.5 & 16.8 & 0.055 & 0.01 & SR & Re\\
069 & J215646.76-065650.3 & ABELL2399 & 0 & -21.64 & 0.31 & 67.3 & 68.0 & 0.7 & 0.23 & 0.009 & FR & Re\\
070 & J215650.44-074111.3 & ABELL2399 & 1 & -20.31 & 0.56 & 154.1 & 148.5 & 5.6 & 0.519 & 0.017 & FR & Re\\
071 & J215653.48-075405.5 & ABELL2399 & 1 & -20.36 & 0.509 & 85.7 & 81.0 & 4.7 & 0.33 & 0.02 & FR & Re\\
072 & J215656.92-065751.3 & ABELL2399 & 0 & -20.43 & 0.17 & 99.0 & 93.0 & 6.0 & 0.196 & 0.017 & FR & Re\\
073 & J215658.25-074910.7 & ABELL2399 & 1 & -20.44 & 0.107 & 67.4 & 21.5 & 45.9 & 0.192 & 0.019 & FR & Re\\
074 & J215658.51-074843.1 & ABELL2399 & 1 & -21.01 & 0.159 & 31.5 & 38.0 & 6.5 & 0.426 & 0.013 & LTG & Re\\
075 & J215701.22-075415.2 & ABELL2399 & 1 & -20.78 & 0.331 & 10.0 & 172.0 & 18.0 & 0.065 & 0.013 & SR & Re\\
076 & J215701.35-074653.3 & ABELL2399 & 1 & -20.65 & 0.215 & 155.1 & 145.5 & 9.6 & 0.193 & 0.014 & FR & Re\\
077 & J215701.71-075022.5 & ABELL2399 & 1 & -22.25 & 0.294 & 103.2 & 118.5 & 15.3 & 0.198 & 0.008 & FR & Re\\
078 & J215705.29-071411.2 & ABELL2399 & 1 & -20.25 & 0.0 & 0.0 & 158.5 & 21.5 & 0 & 0 & - & -\\
079 & J215716.83-075450.5 & ABELL2399 & 1 & -20.86 & 0.294 & 128.7 & 128.0 & 0.7 & 0.271 & 0.013 & FR & Re\\
080 & J215721.41-074846.8 & ABELL2399 & 1 & -20.93 & 0.155 & 91.3 & 100.0 & 8.7 & 0.127 & 0.012 & FR & Re\\
081 & J215723.40-075814.0 & ABELL2399 & 1 & -21.24 & 0.292 & 105.4 & 114.0 & 8.6 & 0.137 & 0.009 & SR & Re\\
082 & J215726.31-075137.7 & ABELL2399 & 1 & -20.35 & 0.277 & 47.3 & 55.5 & 8.2 & 0.34 & 0.021 & FR & Re\\
083 & J215727.30-073357.6 & ABELL2399 & 1 & -20.64 & 0.343 & 158.8 & 167.5 & 8.7 & 0.275 & 0.012 & FR & Re\\
084 & J215727.63-074812.8 & ABELL2399 & 1 & -20.35 & 0.478 & 154.2 & 157.0 & 2.8 & 0.447 & 0.012 & FR & Re\\
085 & J215728.65-073155.4 & ABELL2399 & 1 & -20.46 & 0.118 & 166.9 & 11.5 & 24.6 & 0.306 & 0.016 & LTG & Re\\
086 & J215729.42-074744.5 & ABELL2399 & 1 & -21.74 & 0.449 & 81.3 & 83.5 & 2.2 & 0.256 & 0.008 & FR & Re\\
087 & J215733.30-074420.6 & ABELL2399 & 1 & -20.88 & 0.312 & 24.4 & 136.5 & 67.9 & 0.231 & 0.012 & LTG & Re\\
088 & J215733.47-074739.2 & ABELL2399 & 1 & -21.67 & 0.325 & 155.2 & 151.5 & 3.7 & 0.202 & 0.007 & FR & Re\\
089 & J215733.72-072729.3 & ABELL2399 & 1 & -21.37 & 0.498 & 165.4 & 169.0 & 3.6 & 0.304 & 0.008 & FR & Re\\
090 & J215743.17-072347.5 & ABELL2399 & 1 & -21.13 & 0.109 & 142.8 & 150.5 & 7.7 & 0.314 & 0.009 & FR & Re\\
091 & J215743.24-074545.1 & ABELL2399 & 1 & -21.06 & 0.202 & 108.8 & 136.5 & 27.7 & 0.131 & 0.011 & SR & Re\\
092 & J215745.05-075701.8 & ABELL2399 & 1 & -20.46 & 0.382 & 56.8 & 56.0 & 0.8 & 0.506 & 0.01 & FR & 2Re\\
093 & J215753.00-074419.0 & ABELL2399 & 1 & -21.15 & 0.058 & 179.1 & 12.0 & 12.9 & 0.267 & 0.012 & FR & Re\\
094 & J215759.85-072749.5 & ABELL2399 & 1 & -21.15 & 0.126 & 49.5 & 32.5 & 17.0 & 0.483 & 0.03 & LTG & Re\\
095 & J215806.62-080642.4 & ABELL2399 & 1 & -20.59 & 0.127 & 139.9 & 133.0 & 6.9 & 0.248 & 0.012 & FR & Re\\
096 & J215807.50-075545.4 & ABELL2399 & 1 & -21.31 & 0.388 & 51.0 & 47.0 & 4.0 & 0.29 & 0.012 & FR & Re\\
097 & J215810.04-074801.4 & ABELL2399 & 1 & -20.58 & 0.4 & 121.9 & 117.0 & 4.9 & 0.364 & 0.016 & FR & Re\\
098 & J215811.35-072654.0 & ABELL2399 & 1 & -20.44 & 0.238 & 149.5 & 152.5 & 3.0 & 0.346 & 0.012 & FR & 2Re\\
099 & J215826.28-072154.0 & ABELL2399 & 1 & -21.01 & 0.13 & 73.2 & 65.5 & 7.7 & 0.427 & 0.072 & LTG & Re2\\
100 & J215840.77-074939.8 & ABELL2399 & 0 & -21.05 & 0.32 & 140.9 & 149.5 & 8.6 & 0.239 & 0.017 & FR & Re\\
101 & J215853.98-071531.8* & ABELL2399 & 0 & -21.5 & 0.379 & 11.1 & 177.5 & 13.6 & 0.435 & 0.017 & LTG & Re2\\
102 & J215902.71-073930.0 & ABELL2399 & 1 & -20.28 & 0.52 & 106.5 & 96.0 & 10.5 & 0.358 & 0.019 & FR & Re\\
103 & J215910.35-080431.2 & ABELL2399 & 0 & -20.99 & 0.23 & 96.2 & 131.5 & 35.3 & 0.607 & 0.02 & LTG & Re\\
104 & J215924.41-073442.7 & ABELL2399 & 1 & -20.53 & 0.22 & 109.9 & 138.5 & 28.6 & 0.427 & 0.041 & LTG & Re\\
105 & J215942.63-073028.6 & ABELL2399 & 0 & -20.44 & 0.448 & 71.4 & 89.5 & 18.1 & 0.589 & 0.042 & LTG & Re\\
106 & J215945.43-072312.3 & ABELL2399 & 0 & -21.45 & 0.108 & 44.8 & 24.5 & 20.3 & 0.087 & 0.009 & SR & Re\\
\hline
\end{tabular}
\caption{The 106 galaxies in the SAMI Pilot Survey. The first column shows the galaxy ID, the second the galaxy name. The third column shows the cluster sample the galaxy belongs to, while the fourth shows whether the galaxy was deemed to be a cluster member (1) or not (0). The absolute R-band magnitudes use for sample selection are given in column five. These values come from the NYU-VAGC \citep{Blanton2005}. The sixth and seventh columns give the ellipticity and PA. These values are measured at one effective radius, except for those galaxies whose names are marked with a *, which showed strong contamination from bars at one effective radius. In these objects the ellipticity and PA are calculated in the outskirts of the galaxy, away from the influence of the bar. The kinematic PA is given in column eight, with the misalignment angle in column nine. The tenth and eleventh columns show $\lambda_{R}$ and the error thereon. The twelfth column gives the kinematic classification for the ETGs and the morphological classification for the LTGs. The final column gives the fiducial radius at which $\lambda_{R}$ was measured.}
\label{tab:properties}
\end{table*}


\section[]{Stellar Kinematic Maps for the SAMI}

\begin{figure*}
\centering
\includegraphics[width=17cm]{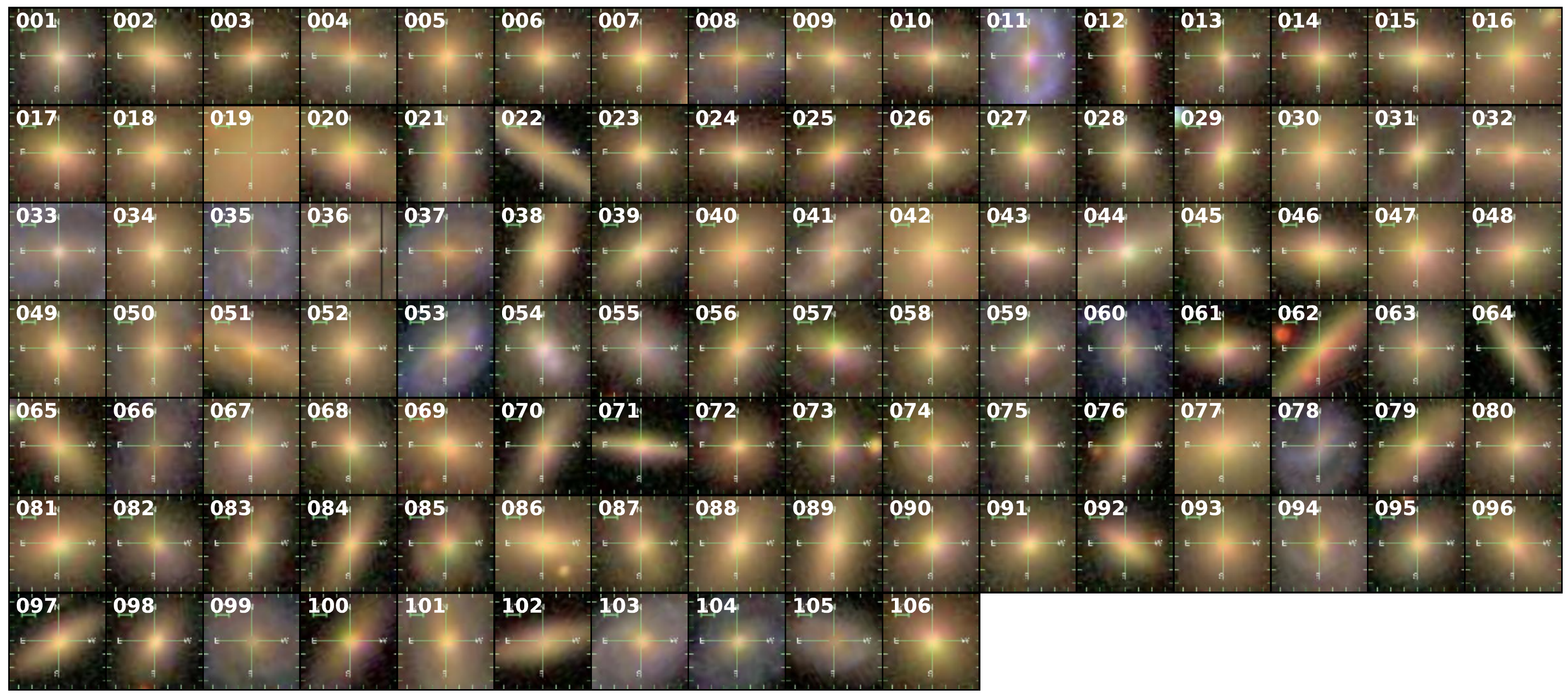}
\caption{The $gri$ SDSS DR7 images for all 106 galaxies in the SAMI Pilot Survey. The galaxy ID shown in the top left of each panel corresponds with that given in Table \ref{tab:properties}.}
\label{fig:sdss_all}
\end{figure*}

\begin{figure*}
\centering
\includegraphics[width=17cm]{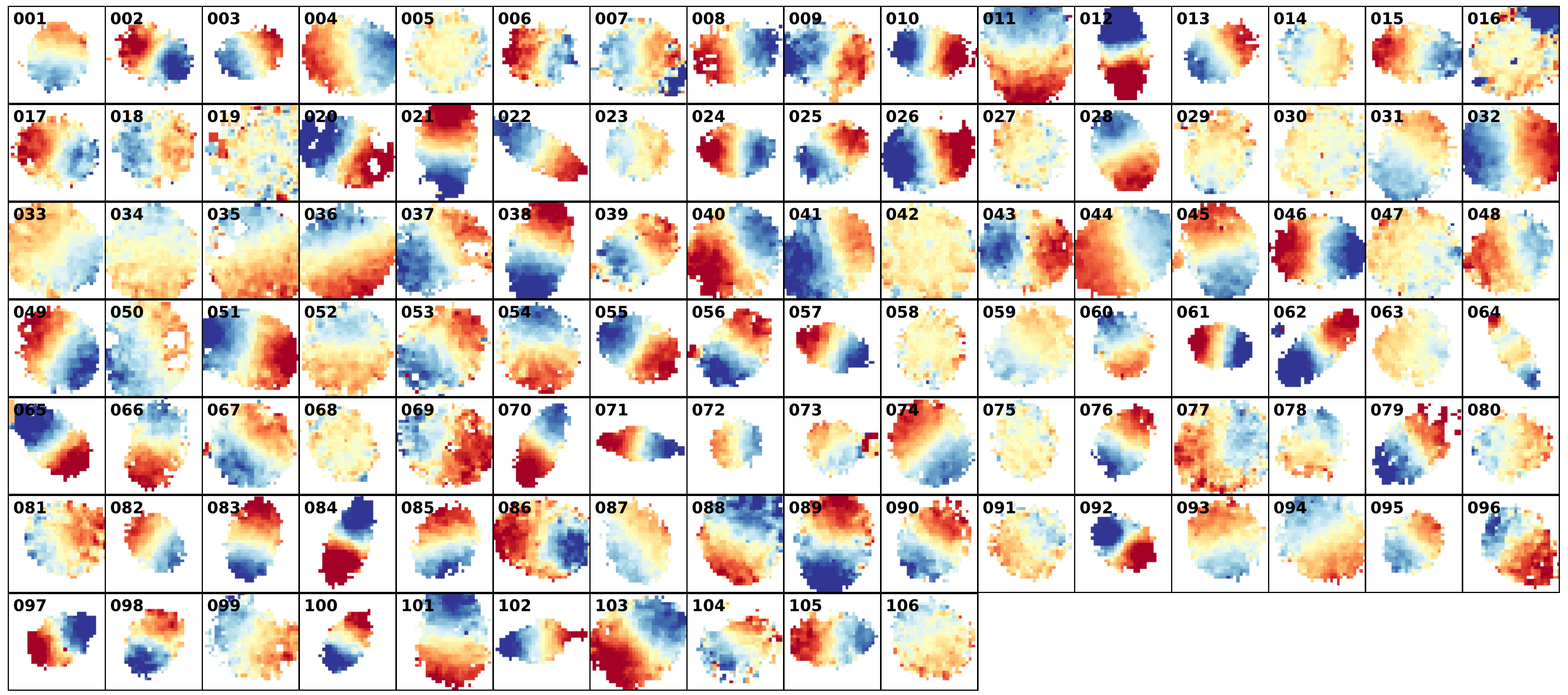}
\caption{The stellar velocity maps for all 106 galaxies in the SAMI Pilot Survey. The galaxy ID shown in the top left of each panel corresponds with that given in Table \ref{tab:properties}. The maps are all scaled to between $\pm150$km\,s$^{-1}$. Each panel is 15 arcseconds on a side.}
\label{fig:vel_all}
\end{figure*}

\begin{figure*}
\centering
\includegraphics[width=17cm]{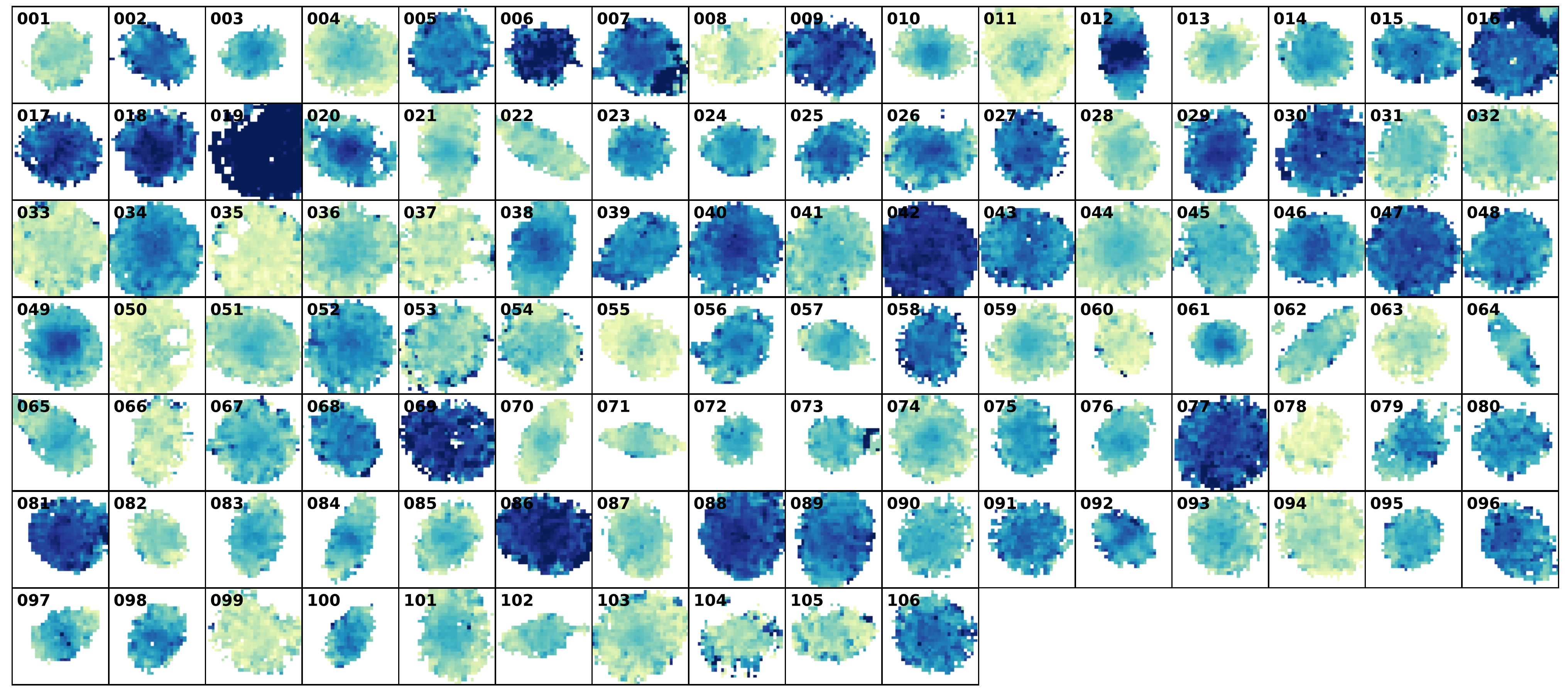}
\caption{The stellar velocity dispersion maps for all 106 galaxies in the SAMI Pilot Survey. The galaxy ID shown in the top left of each panel corresponds with that given in Table \ref{tab:properties}. The maps are all scaled to between $0-300$km\,s$^{-1}$. Each panel is 15 arcseconds on a side.}
\label{fig:sigma_all}
\end{figure*}

The $gri$ SDSS DR7 images of the SAMI Pilot Survey galaxies are shown in Figure \ref{fig:sdss_all}. The kinematic maps, derived as described in Section \ref{sec:der_params} are given in Figures \ref{fig:vel_all} and \ref{fig:sigma_all}.

\end{document}